\documentclass[11pt]{amsart}
\usepackage{amsmath,amssymb}

\newtheorem{teo}{Theorem}[section]
\newcommand{\ir}{{\rm I}\!{\rm R}}

\newtheorem{defi}{Definition}[section]
\newtheorem{lem}{Lemma}[section]

\newtheorem{rema}{Remark}[section]

\begin{document}

\title[Hydrodynamic limit of a BGK model]
{Hydrodynamic limit of a B.G.K. like model on domains with
boundaries and analysis of kinetic boundary 
conditions for scalar multidimensional
conservation laws}

\author{M. Tidriri}

\address{M. Tidriri\\
Department of Mathematics \\
ISU, Ames, IA 50011-2064}
\email{tidriri@iastate.edu}

\date{}



\begin{abstract}
In this paper we study the hydrodynamic limit of a B.G.K. like kinetic
model on domains with boundaries via $BV_{loc}$ theory. We obtain as a
consequence existence results for scalar
multidimensional conservation laws with kinetic boundary conditions. 
We require that the initial
and boundary data satisfy the optimal assumptions that they all belong
to $L^1\cap L^\infty$ with the additional regularity assumptions that
the initial data are in $BV_{loc}$. 
We also extend our hydrodynamic analysis to the case of a generalized
kinetic model to account for forces effects and we obtain as a 
consequence the existence theory for conservation laws with source
terms and kinetic boundary conditions. 
\end{abstract}

\thanks{The author is partially supported by the Air Force Office of 
Scientific Research under Grant F496200210354.
This paper will appear in Journal of Statistical Physics}

\maketitle

%
%
%

\section{Introduction}

In this paper we consider the following kinetic model 

\begin{eqnarray}
& & [\partial_t+a(v)\cdot \partial_x]
g_{\epsilon}(x,v,t)=\frac{1}{\epsilon}
(\chi_{w_{\epsilon}(x,t)}(v)-g_{\epsilon} (x,v,t)) 
\;\; \mbox{in}\;\; \Omega\times V\times (0,T) \label{eq11}\\
& & g_{\epsilon}(x,v,t)=g_{\epsilon 0}(x,v,t) \;\; \mbox{on}\;\;
\Gamma_0^-\times (0,T) \label{eq13}\\
& & g_{\epsilon}(x,v,t)=g_{\epsilon 1}(x,v,t) \;\; \mbox{on}\;\;
\Gamma_1^{-}\times (0,T), \;\; \;\;  \label{eq12}\\
& & \nonumber \\
& & g_\epsilon(x,v,0)=g^0_\epsilon(x,v) 
\;\;\mbox{in}\;\;\Omega\times V\label{eq14}
\end{eqnarray}

\noindent and study its relation to the scalar multidimensional
conservation laws

\begin{eqnarray}
& & \partial_t w+\partial_{x_i} [A_i(w)] =0 
\;\; \mbox{in}\;\; \Omega_{g}\times (0,T) \label{eq3n}\\
& & \mbox{Boundary conditions for $w$ on $\Gamma_0\times (0,T)$
and $\Gamma_1\times (0,T)$}
\label{eq4}\\
& & \nonumber \\
& & w(x,0)=w^0(x) \;\;\mbox{in}\;\; \Omega \label{eq6}
\end{eqnarray}

\noindent Here, $\Omega=(0,1) \times \ir^{d-1}$ is the physical
domain. The boundaries are defined as follows

\begin{eqnarray*}
\Gamma_0&=&\{ 0\}\times\ir^{d-1}, \;\;\; 
\Gamma_1=\{ 1\}\times \ir^{d-1},
\nonumber \\
\Gamma_0^-&=&\{ (x,v)\in \{0\}\times \ir^{d-1} \times V: \;\; a(v)\cdot n(x)
<0\} \nonumber \\
\Gamma_1^-&=&\{ (x,v)\in \{1\}\times \ir^{d-1}\times V:\;\; a(v)\cdot n(x)
<0\} 
\end{eqnarray*}

\noindent where $n$ denotes the exterior unit
normal vector to $\Omega$. 
The boundary conditions in (\ref{eq4}) for the
conservation laws are
prescribed on a part of $\Gamma_0$ resp. $\Gamma_1$.
These boundary conditions will be precised in Definition \ref{defi1}.
The set $V=\ir$ is the velocity domain.
The function $g_\epsilon$ describes the microscopic density of particles 
at $(x,t)$ with velocity $v$ in the kinetic domain. The function $w$
describes the local density of particles at $(x,t)$ in the hydrodynamic
domain. The physical parameter $\epsilon>0$ is the microscopic scale.
The functions 
$g_\epsilon^0$ and $w^0$ are the initial data while 
$g_{\epsilon 0}$ and $g_{\epsilon 1}$ are boundary data. 
The boundary conditions in (\ref{eq4}) involve also $w_1$ and $g_0$
which are given boundary data; see Definition \ref{defi1} below.
Let $A=(A_i)_{1\le i\le d},$
the components of $A$ are assumed to satisfy $A_i(\cdot) \in C^1$ and are
related to $a_i(\cdot)$ by $a_i(\cdot)=A_i^\prime(\cdot)$,
$i=1,\cdots,d$. The local density of
particles $w_\epsilon$ at $(x,t)$ is related to the microscopic density
$g_\epsilon$ by $w_\epsilon(x,t)=\int_V g_\epsilon(x,v,t)dv$. The
collisions in the kinetic domain are given by the nonlinear kernel in
the right hand side of Eq. (\ref{eq11}) in which $\chi_u(v)$ is the
signature of $u$ defined by

\begin{equation}\label{eqsig}
\chi_u(v)=\left \{ \begin{array}{lll}
                   +1 \; & \mbox{if $0<v\le u$} \\
                   -1 \; & \mbox{if $u\le v<0$} \\
                   0             & \mbox{otherwise}
                   \end{array}
                  \right.
\end{equation}

Our main objective in this paper is to describe the conservation laws
(\ref{eq3n})-(\ref{eq6}) as the macroscopic limit of the Boltzmann-like
equations (\ref{eq11})-(\ref{eq14}), as the microscopic scale,
$\epsilon>0$, goes to $0$.
This problem is a particular case of the more general problem of 
describing compressible Euler equations as the macroscopic limit of
Boltzmann or B.G.K. equations, as the microscopic scale goes to $0$.
The convergence of the moments of the kinetic distributions of Boltzmann
or B.G.K. equations to weak solutions of the compressible Euler
equations is still an open problem. In the case of strong solutions this
question has been solved by Caflisch \cite{caflisch}. The case of
domains with boundaries is still completely open.

The study of the hydrodynamic limit of the kinetic model
(\ref{eq11})-(\ref{eq14}) in full space ($\Omega=\ir^d$)
has been performed by Perthame and Tadmor \cite{bp}. They proved that this
model converges as the microscopic scale goes to $0$ to a conservation laws
of the form in (\ref{eq3n}). Later Nouri, Omrane, and Vila attempted to
study this hydrodynamic limit in the case of $\ir^{+}\times\ir^{d-1}$ 
\cite{nouri}. Unfortunately their proofs are wrong.
In their proofs of the various $L^\infty$, $L^1$, and $BV$ uniform, in
$\epsilon$, estimates, they have used in an essential way Gronwall lemma,
which does not yield the uniform bounds they claimed.
These uniform bounds are central to their proofs. Therefore their
proofs are wrong.
In \cite{nouri}, Proposition 3 on page 784 and Proposition 4 on page 
786, are obtained by applying Gronwall lemma to the inequality

\begin{eqnarray}
V_\epsilon(t)\le \int_0^t \frac{1}{\epsilon} e^{(s-t)/\epsilon}
V_\epsilon (s)ds+C \label{gron1}
\end{eqnarray}

\noindent and then they conclude that $|V_\epsilon(t)|$ is uniformly
bounded. This is not the case as the following counterexample shows.
Take $V_\epsilon(t)=\frac{C t}{\epsilon}$. $V_\epsilon$ satisfies the
inequality (\ref{gron1}), however, $V_\epsilon$ is not uniformly in
$\epsilon$ bounded.

In this paper, we shall see how the ideas developed by the author 
in \cite{mdt-bv,mdt-crkfbc} 
to study the more difficult coupled system of kinetic equations (\ref{eq11})
and their hydrodynamic limit (conservation laws of the form in 
(\ref{eq3n})), which is a simplified case of the more general 
coupled system of Boltzmann equations and their hydrodynamic limits
(compressible Euler and Navier-Stokes equations) introduced and
studied in \cite{mdt2,mdt3,mdt4} (see also the references therein),
can be applied to study the hydrodynamic limit of  the kinetic model
(\ref{eq11})-(\ref{eq14}) in the case of domains with boundaries.
Our proofs rely on optimal assumptions on the initial and boundary
data and do not use any technical assumptions. 
For a further study of this problem and a generalization of the concept
of kinetic formulation to conservation laws on domains with boundaries,
we refer to the author's work \cite{mdtkf2}. 

In the second part of this paper, we introduce a generalization of the 
kinetic model (\ref{eq11})-(\ref{eq14}) that includes forces effects and  
whose macroscopic limit, as the microscopic scale go to $0$, yields
conservation laws with source terms. This kinetic model is more
appropriate to describe the physics at the microscopic level than the
model proposed in \cite{bp} for the approximation of conservation laws
with source terms. We then study the hydrodynamic limit of the proposed
kinetic model and prove the existence theory for its continuum limit,
i.e. the conservation laws with source terms.

This paper is organized as follows. In the next section we 
study the kinetic problem. We prove various a priori estimates that
are needed for the study of the hydrodynamic limit of the kinetic
problem. 
In Section 3, we precise our definition of physically correct solution to 
the problem (\ref{eq3n})-(\ref{eq6}). We then study the hydrodynamic limit
of the kinetic
problem and prove our main result. 
Finally in Section 4, we study the one dimentional case via compensated 
compactness. We prove the convergence of the moments of the
kinetic distributions to the solution of the conservation laws
without any compactness argument (based on
$BV_{loc}$ theory).

\section{The kinetic equations}

In this section we shall study various properties of the solution of the 
kinetic equations (\ref{eq11})-(\ref{eq14}).
Some of our proofs are closely related to those for the full space case 
in \cite{bp}. However, our
problem is on a domain with boundaries. 
This introduces new difficulties that are not present in the
full space case.
These difficulties must be handled by different techniques.
We begin by stating a result about the well posedness of the kinetic
problem (\ref{eq11})-(\ref{eq14}). We
then establish various properties of the solution, including 
$L^\infty$, $L^1$, and $BV_{loc}$ estimates. 
These estimates will be used for the study of the hydrodynamic limit
of Problem (\ref{eq11})-(\ref{eq14}) as $\epsilon \rightarrow 0$.
We shall also use the following notations. 

\begin{eqnarray*}
\Omega_0&=& \{ (x,v,t)\in \Omega \times V\times (0,T):
                       x_1-a_1(v)t <0 \} \\
\Omega_{01}&=& \{ (x,v,t)\in \Omega \times V\times (0,T):
                       0<x_1-a_1(v)t <1 \} \\
\Omega_{1}&=& \{ (x,v,t)\in \Omega \times V\times (0,T):
                       x_1-a_1(v)t >1 \} 
\end{eqnarray*}

\noindent where $x=(x_1,x_\star)$.

\subsection{Existence theory and basic estimates}

We establish in this section the existence and
uniqueness theory and derive basic estimates for the solutions
of the kinetic equations.

\begin{teo}\label{teok}
Assume that

\[
g^0_\epsilon \in L^1(\Omega\times V), \;\;
a(v)\cdot n g_{\epsilon 1}\in 
L^1(\Gamma_1^{-}\times (0,T)), \;\;
a(v)\cdot n g_{\epsilon 0} \in L^1(\Gamma_0^{-}\times (0,T))
\]

\noindent
Then the problem (\ref{eq11})-(\ref{eq14}) has a unique solution
$g_\epsilon$ in
$L^\infty((0,T);L^1(\Omega \times V)).$  Moreover,
$g_\epsilon$ satisfies the integral representation

\begin{eqnarray*}
& & \mbox{In} \;\; \Omega_0\;\; \\ 
& &
\;\;g_\epsilon(x,v,t)=g_{\epsilon 0}(x_\star -\frac{x_1}{a_1(v)}a_\star(v),
v,t-\frac{x_1}{a_1(v)})\mbox{exp}({-x_1/(a_1(v)\epsilon)})
\\ & &
+\frac{1}{\epsilon} \int_{t-\frac{x_1}{a_1(v)}}^t
e^{(s-t)/\epsilon}\chi_{w_\epsilon(x(s),s)}(v)ds \\
& & \\
& & \mbox{In} \;\; \Omega_{01}\\
& &
\;\; g_\epsilon(x,v,t)=g^0_\epsilon(x-a(v)t,v)\mbox{exp}(-t/\epsilon)
+\frac{1}{\epsilon} \int_{0}^t
e^{(s-t)/\epsilon}\chi_{w_\epsilon(x(s),s)}(v)ds \\
& & \\
& & \mbox{In} \;\; \Omega_1 \;\; \\ 
& &
\;\; g_\epsilon(x,v,t)=g_{\epsilon 1}
(x_\star+\frac{1-x_1}{a_1(v)}a_\star(v),v,t-\frac{x_1-1}{a_1(v)})
\mbox{exp}{((1-x_1)/\epsilon a_1(v))} 
\\ & &
+\frac{1}{\epsilon} \int_{t-\frac{x_1-1}{a_1(v)}}^t
e^{(s-t)/\epsilon}\chi_{w_\epsilon(x(s),s)}(v)ds \\
\end{eqnarray*}

\noindent where $x(s)=x+(s-t)a(v)$, $x=(x_1,x_\star)$, and
$a(v)=(a_1(v),a_\star(v))$.

Finally, let $g_\epsilon$ and
$G_\epsilon$ be two solutions of (\ref{eq11})-(\ref{eq14})
with corresponding densities $w_\epsilon(x,t)=\int_V g_\epsilon(x,v,t)dv$ and
$W_\epsilon(x,t)=\int_V G_\epsilon(x,v,t)dv$; and let 
$g^0_\epsilon,\;g_{\epsilon 0},\;g_{\epsilon 1}$ 
resp. $G^0_\epsilon,\;G_{\epsilon 0},\;G_{\epsilon 1}$ 
denote the corresponding data. We have

\begin{eqnarray}
& &
\|g_\epsilon-G_\epsilon\|_{L^1(\Omega\times V)}
+\|a(v)\cdot n (g_{\epsilon}-G_{\epsilon})\|_
{L^1(\Gamma_0^{+}\times (0,t))}+
\nonumber \\
& &
\|a(v)\cdot n (g_\epsilon-G_\epsilon)\|_{L^1(\Gamma_1^{+}\times (0,t))}
\nonumber \\
&\le& 
\|g_\epsilon^0-G_\epsilon^0\|_{L^1(\Omega \times V)}
+\|a(v)\cdot n (g_{\epsilon 0}-G_{\epsilon
0})\|_{L^1(\Gamma_0^{-}\times(0,t))}+
\nonumber \\
& & 
\|a(v)\cdot n (g_{\epsilon 1}-G_{\epsilon
1})\|_{L^1(\Gamma_1^{-}\times (0,t))}
\label{estcont}
\end{eqnarray}

\end{teo}

\begin{rema}
Although we can derive contraction properties directly from the integral
representation, we prefer to use a different method, which allows us to
obtain the inequalities in (\ref{estcont}).
\end{rema}

\vskip .5cm
\noindent
{\bf Proof of Theorem \ref{teok}}
\vskip .5cm

We begin with proving the uniqueness and the continuous
dependence of the solution on the data given in (\ref{estcont}). 
These estimates are needed for the proofs of various results below.
Therefore, we shall give a somewhat detailed proof.
The idea of the proof is to use a combination of the author's method
\cite{mdt1,mdt2} and ideas from \cite{krushkov}.

The function $G_\epsilon$ satisfies an equation similar to
Eq. (\ref{eq11}). Subtracting this equation from Eq. (\ref{eq11}),
and multiplying the resulting equation by $\varphi$ 
a test function in
$C^1(\bar{\Omega}\times V\times [0,T])$ to be precised later, 
and integrating by parts, we obtain

\begin{eqnarray}
& &
\int_{\Omega \times V}((g_\epsilon-G_\epsilon)\varphi)(\cdot,\cdot,t)-
\int_{\Omega \times V}((g_\epsilon-G_\epsilon)\varphi)(\cdot,\cdot,0)
\nonumber \\
& &
-\int_{\Omega \times V\times (0,t)}
(\partial_t+a(v)\cdot \partial_x)(\varphi)(g_\epsilon-G_\epsilon)
\nonumber \\
& & 
+\int_{\Gamma_0^-\times (0,t)} a(v)\cdot n (g_{\epsilon 0}-g_{\epsilon 0})
\varphi 
+\int_{\Gamma_1^-\times (0,t)} a(v)\cdot n (g_{\epsilon 1}-G_{\epsilon 1}) 
\varphi \nonumber \\
& & 
+\int_{\Gamma_0^+\times (0,t)} a(v)\cdot n (g_{\epsilon}-G_{\epsilon})
\varphi 
+\int_{\Gamma_1^+\times (0,t)} a(v)\cdot n (g_{\epsilon}-G_{\epsilon}) 
\varphi \nonumber \\
&=& \frac{1}{\epsilon}\int_{\Omega\times V\times (0,t)}
((\chi_{w_\epsilon}-\chi_{W_\epsilon})-(g_\epsilon-G_\epsilon))\varphi
\label{ke1}
\end{eqnarray}

We then take $\varphi=\mbox{sign}^{\mu}
(g_\epsilon-G_\epsilon)\psi(x,t)$
with $x\mbox{sign}^{\mu} (x) \ge 0 \;\; x\in \ir$,
and $\psi$ is a nonnegative test function
and $\mbox{sign}^{\mu}$ is a regularization of $\mbox{sign}$
function. 
Plugging in (\ref{ke1}) and passing to the limit as $\mu \rightarrow 0$,
we obtain

\begin{eqnarray}
& &
\int_{\Omega \times V}(|g_\epsilon-G_\epsilon|\psi)(\cdot,\cdot,t)
+\int_{\Gamma_0^+\times (0,t)} a(v)\cdot n |g_\epsilon-G_\epsilon|
\psi \nonumber \\
& &
+\int_{\Gamma_1^+\times (0,t)} a(v)\cdot n |g_\epsilon-G_\epsilon| \psi 
+\int_{\Gamma_0^{-}\times (0,t)} a(v)\cdot n |g_{\epsilon
0}-G_{\epsilon 0}|\psi +\nonumber \\
& &
\int_{\Gamma_1^-\times (0,t)} a(v)\cdot n |g_{\epsilon 1}-G_{\epsilon
1}| \psi -
\int_{\Omega \times V}(|g_\epsilon-G_\epsilon|\psi)(\cdot,\cdot,0)
\nonumber \\
&=& 
\frac{1}{\epsilon}\int_{\Omega\times V\times (0,t)}
[(\chi_{w_\epsilon}-\chi_{W_\epsilon})
\mbox{sign}(g_\epsilon-G_\epsilon)-|g_\epsilon-G_\epsilon|]\psi+
\nonumber \\
& &
\int_{\Omega \times V\times (0,t)}
(\partial_t+a(v)\cdot \partial_x)(\psi) |g_\epsilon-G_\epsilon|
\label{eqmul1}
\end{eqnarray}

Using the properties of $\chi$, this yields

\begin{eqnarray}
& &
\int_{\Omega \times V}(|g_\epsilon-G_\epsilon|\psi)(\cdot,\cdot,t)
+\int_{\Gamma_0^+\times (0,t)} a(v)\cdot n |g_\epsilon-G_\epsilon|
\psi \nonumber \\
& &
+\int_{\Gamma_1^+\times (0,t)} a(v)\cdot n |g_\epsilon-G_\epsilon| \psi 
\nonumber \\
&\le & 
\int_{\Omega \times V}(|g_\epsilon-G_\epsilon|\psi)(\cdot,\cdot,0)
-\int_{\Gamma_0^{-}\times (0,t)} a(v)\cdot n |g_{\epsilon
0}-G_{\epsilon 0}|\psi \nonumber \\
& &
-\int_{\Gamma_1^-\times (0,t)} a(v)\cdot n |g_{\epsilon 1}-G_{\epsilon
1}| \psi +
\int_{\Omega \times V\times (0,t)}
(\partial_t+a(v)\cdot \partial_x)(\psi) |g_\epsilon-G_\epsilon|
\label{md1}
\end{eqnarray}

\noindent Taking now $\psi(t)\equiv 1 $ yields the estimate
(\ref{estcont}).

To prove the existence of a solution to the kinetic problem, we use the
following iterations 

\begin{eqnarray}
& & [\partial_t+a(v)\cdot \partial_x]
g_{\epsilon}^{n+1}(x,v,t)=\frac{1}{\epsilon}
(\chi_{w_{\epsilon}^{n}(x,t)}(v)-g_{\epsilon}^{n+1} (x,v,t)) 
\;\; \mbox{in}\;\; \Omega\times V\times (0,T) \label{it1}\\
& & g_{\epsilon}^{n+1}(x,v,t)=
g_{\epsilon 0}(x,v,t) \;\; \mbox{on}\;\;
\Gamma_0^-\times (0,T) \label{it2}\\
& & g_{\epsilon}^{n+1}(x,v,t)=g_{\epsilon 1}(x,v,t) \;\; \mbox{on}\;\;
\Gamma_1^{-}\times (0,T), \;\; \;\;  \label{it3}\\
& & \nonumber \nonumber \\
& & g_\epsilon^{n+1}(x,v,0)=g^0_\epsilon(x,v) \;\;\mbox{in}\;\;
\Omega \times V. \label{it4}
\end{eqnarray}

Using (\ref{eqmul1}) in the present context with
$g_\epsilon=g^{n+1}_\epsilon$ and $G_\epsilon=g^{m+1}_\epsilon$, and 
using the properties of $\chi$, we obtain

\begin{eqnarray}
& &
\int_{\Omega \times V}(|g_\epsilon-G_\epsilon|\psi)(\cdot,\cdot,t)
+\int_{\Gamma_0^+\times (0,t)} a(v)\cdot n |g_\epsilon-G_\epsilon|
\psi+ \nonumber \\
& &
\int_{\Gamma_1^+\times (0,t)} a(v)\cdot n |g_\epsilon-G_\epsilon| 
\psi +
\frac{1}{\epsilon}\int_{\Omega\times V\times (0,t)}
|g_\epsilon-G_\epsilon|\psi\nonumber \\
&=& 
\frac{1}{\epsilon}\int_{\Omega\times V\times (0,t)}
(\chi_{w_\epsilon^{n}}-\chi_{W_\epsilon^{m}})
\mbox{sign}(g_\epsilon-G_\epsilon)\psi+ \nonumber \\
& & 
\int_{\Omega \times V\times (0,t)}
(\partial_t+a(v)\cdot \partial_x)(\psi)|g_\epsilon-G_\epsilon| 
\nonumber \\
&\le & 
\int_{\Omega \times V\times (0,t)}
(\partial_t+a(v)\cdot \partial_x)(\psi)|g_\epsilon-G_\epsilon| 
+\frac{1}{\epsilon}\int_{\Omega\times V\times (0,t)}
|g_\epsilon^{n}-g_\epsilon^{m}|\psi \nonumber \\
\label{eqmul2}
\end{eqnarray}

\noindent Taking $\psi=e^{-\frac{\alpha}{\epsilon}s}$, $0\le
s\le t$, with $\alpha$ a positive constant, we then obtain

\begin{eqnarray}
& &
\int_{\Omega \times V}(|g_\epsilon^{n+1}-g_\epsilon^{m+1}|\psi)(\cdot,\cdot,t)
+\int_{\Gamma_0^+\times (0,t)} a(v)\cdot n 
|g_\epsilon^{n+1}-g_\epsilon^{m+1}|\psi \nonumber \\
& &
+\int_{\Gamma_1^+\times (0,t)} a(v)\cdot n |g_\epsilon^{n+1}-g_\epsilon^{m+1}| 
\psi +
\frac{1+\alpha}{\epsilon}\int_{\Omega\times V\times (0,t)}
|g_\epsilon^{n+1}-g_\epsilon^{m+1}|\psi\nonumber \\
&\le& 
\frac{1}{\epsilon}\int_{\Omega\times V\times (0,t)}
|g_\epsilon^{n}-g_\epsilon^{m}|\psi
\label{eqmul4}
\end{eqnarray}

Hence we obtain

\begin{eqnarray}
\int_{\Omega\times V\times (0,t)}\psi
|g_\epsilon^{n+1}-g_\epsilon^{m+1}|
&\le &
\frac{1}{1+\alpha}\int_{\Omega\times V\times (0,t)}
|g_\epsilon^n-g_\epsilon^m| \psi\label{eqmul5}
\end{eqnarray}

This and a reuse of (\ref{eqmul4}) proves that the iterations
are contracted to the unique fixed point 
in $L^\infty([0,T];L^1(\Omega\times V))$, which 
satisfies Eq. (\ref{eq11}) and also
the boundary and initial conditions 
(\ref{eq13})--(\ref{eq14})
We also infer from the 
inequality (\ref{estcont}) that the solution $g_\epsilon$ 
depends continuously on the initial and boundary data. 

The integral representation is obtained using the characteristic method.
The proof of the theorem is now finished.

\subsection{Kinetic entropy}

We shall prove an entropy inequality for the solution of the kinetic
problem. This is stated in the following theorem.

\begin{teo} \label{teoent}
The solution to the kinetic problem satisfies the
relation

\begin{eqnarray}
& &
-\int_{\Omega \times V\times (0,T)}
(\partial_t+a(v)\cdot \partial_x)(\psi)|g_\epsilon-\chi_k|
+\int_{\Gamma_0^-\times (0,T)} a(v)\cdot n 
\psi|g_{\epsilon 0}-\chi_k|+ \nonumber \\
& &
\int_{\Gamma_1^{-}\times (0,T)} a(v)\cdot n 
\psi|g_{\epsilon 1}-\chi_k|
\le 0 
\label{ke4}
\end{eqnarray}

\begin{eqnarray*}
& & \forall \psi \in C_0^1(\bar{\Omega}\times V\times (0,T)), \;
\psi \ge 0,\; \forall k\in \ir 
\end{eqnarray*}

\end{teo}

\vskip .5cm
{\bf Proof}
\vskip .5cm

Multiplying Eq. (\ref{eq11}) by 
$\varphi=\mbox{sign}^{\mu} (g_\epsilon-\chi_k)\psi(x,t)$
with $\mbox{sign}^{\mu} (x)$ the regularization of $\mbox{sign}$ function
mentioned in the proof of Theorem \ref{teok}, and $\psi$
is a nonnegative test function in 
$C_0^1(\bar{\Omega}\times V \times (0,T))$, 
and proceeding as in the proof of 
Theorem \ref{teok}, and using the properties of $\chi_w$ the desired
entropy inequality of the theorem.

\subsection{Basic estimates of the solution}

We shall state and prove here some basic estimates for the solution of
the kinetic problem.
We begin with $L^\infty$ estimates.

\begin{lem}\label{lem1}
Assume that 
\[
\|g_{\epsilon 0}\|_{L^\infty(\Gamma_0^{-}\times [0,T])} <C_1,\; 
\|g^0_\epsilon\|_{L^\infty(\Omega\times V)}<C_2, \; 
\|g_{\epsilon 1}\|_{L^\infty(\Gamma_1^{-}\times [0,T])}<C_3
\]

\noindent with $C_1,C_2,$ and $C_3$ positive constants independent of 
$\epsilon$.  Then 
$g_\epsilon$ is uniformly bounded in
$L^\infty(\Omega \times V\times[0,T])$.
Moreover we have

\begin{eqnarray*}
\|g_\epsilon\|_{\infty}  
&\le&  
\mbox{max}(\|g_{\epsilon 0}\|_{L^\infty(\Gamma_0^{-}\times [0,T])},
\|g^0_\epsilon\|_{L^\infty(\Omega\times V)},
\|g_{\epsilon 1}\|_{L^\infty(\Gamma_1^{-}\times [0,T])})+1
\end{eqnarray*}
\end{lem}

\vskip .5cm
\noindent
{\bf Proof:} The proof is based on the use of 
the integral representation of the solution respectively on
$\Omega_0$, 
$\Omega_{01}$, and 
$\Omega_{1}$.

\vskip .5cm
We now present estimates of $g_\epsilon$ and $w_\epsilon$ in 
$L^\infty([0,T];L^1(\Omega\times V))$
and $L^\infty([0,T];L^1(\Omega))$ respectively.

\begin{lem}\label{lem2}

\noindent
Assume that

\begin{eqnarray*}
& & \|a(v)\cdot n g_{\epsilon 0}\|_{L^1(\Gamma_0^{-}\times (0,T))} <C_1,\;\;
\|g^0_\epsilon\|_{L^1(\Omega\times V)}<C_2, \\
& &
\|a(v)\cdot n g_{\epsilon 1}\|_{L^1(\Gamma_1^{-}\times (0,T))}<C_3  
\end{eqnarray*}

\noindent
with $C_1,C_2,$ and $C_3$ positive constants independent of $\epsilon$.
Then $g_\epsilon$ is uniformly bounded in
$L^\infty([0,T];L^1(\Omega\times V))$ and
$w_\epsilon$ is uniformly bounded  in 
$L^\infty([0,T];L^1(\Omega))$. Moreover, we have

\begin{eqnarray*}
\|w_\epsilon\|_{L^\infty([0,T];L^1(\Omega))} 
&\le& 
\|g_\epsilon\|_{L^\infty([0,T];L^1(\Omega\times V))}
\nonumber \\
&\le &
\|a(v)\cdot n g_{\epsilon 0}\|_{L^1(\Gamma_0^{-}\times (0,T))}+
\|a(v)\cdot n g_{\epsilon 1}\|_{L^1(\Gamma_1^{-}\times (0,T))}
\nonumber \\
& &
+\|g_\epsilon^0\|_{L^1(\Omega\times V)}
\end{eqnarray*}

\end{lem}

\vskip .5cm
\noindent
{\bf Proof:}
 Using Formula (\ref{estcont}) with $G_\epsilon\equiv 0$, we obtain

\begin{eqnarray*}
\int_{\Omega\times V}|g_\epsilon(x,v,t)|
&\le& \int_{\Omega \times V}
|g_\epsilon^0(x,v)| + 
\int_{\Gamma_0^{-}\times (0,T)}
|a(v)\cdot n g_{\epsilon 0}|+
\int_{\Gamma_1^{-}\times (0,T)}
|a(v)\cdot n g_{\epsilon 1}|
\end{eqnarray*}

The lemma then follows.

\vskip .5cm
Next we shall show that under the conditions that the
supports in $v\in V$ of the data are compact, the supports in $v\in V$
of $g_\epsilon$ remain compactly
supported with supports included in a fixed compact set independent of
$\epsilon$. We shall also
give some information about the speed of propagation
$a(v)$. This is stated in the following lemma.

\begin{lem} \label{lemc1}
Assume that 
\begin{eqnarray*}
& &
\|g_{\epsilon 0}\|_{L^\infty(\Gamma_0^{-}\times [0,T])}<C_1,\;\;
\|g^0_\epsilon\|_{L^\infty(\Omega\times V)}<C_2,\\
& &
\|g_{\epsilon 1}\|_{L^\infty(\Gamma_1^{-}\times [0,T])}<C_3
\end{eqnarray*}

\noindent with $C_1,C_2,$ and $C_3$ positive constants 
independent of $\epsilon$. 
Assume also that the initial and boundary data 
$g_\epsilon^0$, $g_{\epsilon 0}$, and $g_{\epsilon 1}$ are compactly 
supported in $v\in V$ with supports included in a fixed compact set 
independent of $\epsilon$. 
Then 

\noindent (i) $w_\epsilon$ is uniformly bounded in
$L^\infty(\Omega\times [0,T])$. 

\noindent (ii)
$g_\epsilon$ remains compactly supported in $v\in V$
with support included in a fixed compact set independent of
$\epsilon$.

\noindent (iii)
The speed of propagation $a(v)$ is finite.
\end{lem}

\begin{rema}\label{linf}
In \cite{bp} the uniform $L^\infty$ boundedness (in $\epsilon$)
of the macroscopic density $u_\epsilon=\int_V f_\epsilon(x,v,t)dv$ 
and hence the compactness 
of the support in $v$ of $f_\epsilon(t,x,v)$ together with 
the finite speed of propagation remained unproven. Since in their proof,
which is given on page 504 lines 6 through 12 of \cite{bp}, 
their argument is wrong. Following we quote lines 6 through 12 of page 504 
of \cite{bp}
{\sf
``2. Finite speed of propagation. We assume that initially,
$f_\epsilon(x,\cdot,0)$ has a compact support in $\ir_v$. Let us first
show that $f_\epsilon(x,\cdot,t)$ remains compactly supported. Indeed,
by (2.6), $f_\epsilon(x,v,t)$ and hence $u_\epsilon(\cdot,t)$ are
uniformly bounded, and therefore the contributions of
$\chi_{u(\cdot,\cdot)}(v)$ on the right hand side of (2.2) are supported
by $v\in [-u_\infty,u_\infty]$, where
$u_\infty=\|u_\epsilon(x,t)\|_{L^\infty(\ir_x^d\times \ir_t^{+})}$. 
Consequently, $f_\epsilon(x,\cdot,t)$ given in 2.2 remains compactly
supported for all $t>0$, with support contained in $\mbox{supp}_v
f_\epsilon(x,\cdot,0)\cup [-u_\infty,u_\infty]\cdots$''}

The argument:  Indeed,
by (2.6), $f_\epsilon(x,v,t)$ and hence $u_\epsilon(\cdot,t)$ are
uniformly bounded, is wrong since the uniform (in $\epsilon$) boundedness of
a function (here $f_\epsilon(x,v,t)$)
in $L^\infty(\ir^d \times \ir\times \ir^+)$ does not in
general yield the uniform boundedness (in $\epsilon$) 
of its velocity average (here 
$u_\epsilon(x,t)=\int_{\ir}f_\epsilon(x,v,t)dv$). Take 
for example the function $h_\epsilon(x,v,t)=e^{-\epsilon |v|}\mbox{exp}
({-t-\sum |x_i|})$ and its velocity average $u_\epsilon(x,t)=
\frac{2}{\epsilon} \mbox{exp}({-t-\sum |x_i|})$. 
 
In \cite{bp1}, in order to obtain 
the uniform (in $\epsilon$) bound of
$u_\epsilon(x,t)=\int_{\ir}f_\epsilon(x,v,t)dv$ in $L^\infty(\ir^d
\times \ir^+)$, and hence to fill the gap of \cite{bp},
the author assumed an additional assumption on the sign
of the data: $f_\epsilon(\cdot,v,0)\mbox{sign}(v)\ge 0$. 
This assumption is quite
restrictive if one wants to study the hydrodynamic limit of the kinetic
model, which was one of the main objectives of the paper \cite{bp}.

Because of the above it is clear that the general proof of the above results
remained open despite the various attempts by various authors. We shall
give below two different proofs. One is general and does not use any
additional assumptions, thus solves also the gap in \cite{bp}, and
the second relies on the additional assumption on the sign of the data, 
and thus allows us to compare the two proofs.

\end{rema}

\vskip .5cm
\noindent
{\bf Proof:} 
\vskip .5cm

(i) 
{\it First and general proof of the uniform in $\epsilon \;\;L^\infty$
bound}
\vskip .3cm

We first notice that for every fixed $\epsilon$, using Gronwall lemma 
we conclude that $g_\epsilon$ is in $L^\infty(\Omega \times (0,T);
L^1(V))$ and hence $w_\epsilon$ is in $L^\infty(\Omega \times (0,T))$.
Observe that such argument does not provide a uniform in $\epsilon$ 
bound of $g_\epsilon$ in $L^\infty(\Omega \times (0,T);
L^1(V))$.

Next we prove that $g_\epsilon$ is uniformly in $\epsilon$ bounded in 
$L^\infty(\Omega \times (0,T); L^1(V))$. We write
the integral representation in $\Omega_0$ in the form

\begin{eqnarray*}
& &
g_\epsilon(x,v,t)=g_{\epsilon 0}(x_\star -\frac{x_1}{a_1(v)}a_\star(v),
v,t-\frac{x_1}{a_1(v)})\mbox{exp}({-x_1/(a_1(v)\epsilon)})
\\ & &
+(1-\mbox{exp}({-x_1/(a_1(v)\epsilon)}))
\frac{\int_{t-\frac{x_1}{a_1(v)}}^t
e^{(s-t)/\epsilon}\chi_{w_\epsilon(x(s),s)}(v)ds}
{\int_{t-\frac{x_1}{a_1(v)}}^t
e^{(s-t)/\epsilon}ds}
\end{eqnarray*}

Thus $g_\epsilon(x,v,t)$ is expressed as a convex combination. So by
Jensen inequality, we obtain for any convex function
$\varphi(g_\epsilon)$,

\begin{eqnarray*}
& &
\varphi(g_\epsilon(x,v,t))
\le \varphi(g_{\epsilon 0}(x_\star -\frac{x_1}{a_1(v)}a_\star(v),
v,t-\frac{x_1}{a_1(v)}))\mbox{exp}({-x_1/(a_1(v)\epsilon)})
\\ & &
+\frac{1}{\epsilon}\int_{t-\frac{x_1}{a_1(v)}}^t
e^{(s-t)/\epsilon}\varphi(\chi_{w_\epsilon(x(s),s)}(v))ds
\end{eqnarray*}

We obtain similar formula for $g_\epsilon(x,v,t)$ in $\Omega_{01}$
and $\Omega_1$. Now taking $\varphi(g)=|g|^p$ and integrating over $x$
and $t$, we obtain

\begin{eqnarray*}
& &
\int_{\Omega\times (0,T)}
|g_\epsilon(x,v,t)|^p dx dt  
\le \int_{\Gamma_0\times (0,T)}
|g_{\epsilon 0}(y,v,t)|^p dydt
+\int_{\Omega}|g_{\epsilon}^0(x,v))|^pdx \\
& & +\int_{\Gamma_1\times (0,T)}|g_{\epsilon 1}(y,v,t)|^pdydt
+\int_{\Omega\times (0,T)}
\frac{1}{\epsilon}\int_0^t e^{(s-t)/\epsilon}
|\chi_{w_\epsilon(x,s)}|dx ds dt
\end{eqnarray*}

Taking the $p-$root of both sides and integrating over $V$, we obtain

\begin{eqnarray}
& &
\int_{V}
(\int_{\Omega\times (0,T)}
|g_\epsilon(x,v,t)|^p dx dt  )^{1/p}dv \nonumber \\
&\le & 4^{1/p}
\mbox{max} [ \int_V(\int_{\Gamma_0\times (0,T)}
|g_{\epsilon 0}(x,v,t)|^p)^{1/p}dv,
\int_V (\int_{\Omega}|g_{\epsilon}^0(x,v))|^pdx)^{1/p}dv,\nonumber \\
& & \int_V 
(\int_{\Gamma_1\times (0,T)}|g_{\epsilon 1}(x,v,t)|^pdx dv)^{1/p}dv,
\int_V  (\int_{\Omega\times (0,T)}
\frac{1}{\epsilon}\int_0^t e^{(s-t)/\epsilon}
|\chi_{w_\epsilon(x,s)}|dx ds dt)^{1/p}dv] \nonumber \\
\label{proot0}
\end{eqnarray}

We only need to prove that $$\int_V (\int_{\Omega\times (0,T)}
\frac{1}{\epsilon}\int_0^t e^{(s-t)/\epsilon}
|\chi_{w_\epsilon(x,s)}|dx ds dt)^{1/p}dv$$ is bounded 
uniformly in $\epsilon$
for $p$ large. The other terms are clearly bounded uniformly in $\epsilon$
for $p$ large. For example, the term 
$\int_V (\int_{\Gamma_0\times (0,T)}
|g_{\epsilon 0}(x,v,t)|^p)^{1/p}dv$ is uniformly bounded 
since by assumption $g_{\epsilon 0}$ is uniformly bounded in $\epsilon$
in $L^\infty(\Gamma_0 \times (0,T)\times L^1(V))$ and similarly for the
other terms.

Let 
\begin{eqnarray}
A_\epsilon&=&\{ (x,v,t)\in \Omega\times V\times (0,T)|\;\; 
|w_\epsilon(x,t)|>|v| \} \nonumber \\
V_{\epsilon}&=&\{ v\in V| \;(x,v,t)\in A_\epsilon \; \mbox{for some
$(x,t)\in \Omega\times (0,T)$}\} \nonumber 
\end{eqnarray}

Let $m_\epsilon$ and $n_{\epsilon}$ denote the Lebesgue measure of
$E_\epsilon$ respectively $V_{\epsilon}$. We know from Lemma \ref{lem2} that 

\begin{eqnarray}
m_\epsilon&=&\int_{\Omega\times V\times (0,T)}
|\chi_{w_\epsilon(x,t)}(v)|dx dv dt =
\int_{\Omega\times (0,T)}
|w_\epsilon(x,t)|dx  dt<C
\label{eqmepsilon}
\end{eqnarray}

\noindent where $C$ is independent of $\epsilon$.

Let $C_0>0$ be a fixed constant.
Let $\Upsilon$ denote the set of all $\epsilon>0$ such that

\begin{equation}
\|w_\epsilon\|_\infty >C_0
\label{eqepsilon}
\end{equation}

We know from the begining of this proof that $w_\epsilon$ is in
$L^\infty(\Omega\times (0,T))$ for every fixed $\epsilon$.
If the set $\Upsilon$ is empty or finite then the proof will be
concluded easily. 
Therefore, we assume that $\Upsilon$ is neither empty nor finite.

We prove the following statements. 

\begin{eqnarray}
& & \exists \beta \;\mbox{with}\; 0<\beta<C_0,\;
\exists E\subset \Omega \times (0,T)\;\mbox{with}\; |E|>0\;
\mbox{such that}\;  \|w_{\epsilon}\|_{\infty,E}>\beta 
\nonumber \\
& & 
\mbox{uniformly in }\;\epsilon \in \Upsilon 
\label{eqc0} \\
& & \exists \gamma>0 \;
\mbox{such that}\;  \gamma <m_\epsilon \;
\mbox{uniformly in }\;\epsilon \in \Upsilon
\label{eqgamma}
\end{eqnarray}

\noindent 
Above 
$|F|$ denotes the Lebesgue measure of
the set $F$. If the set $E$ is of infinite measure, then
any subset $E^\prime$ of $E$ satisfying $0<|E^\prime|<\infty$ 
is enough for our purpose. So we may assume that the set $E$ in
(\ref{eqc0}) satisfies $0<|E|<\infty$. This is important 
since we will use below Egoroff theorem for sequence defined on 
such set $E$.

We proceed now to prove (\ref{eqc0}) and (\ref{eqgamma}).
If (\ref{eqc0}) is not true then 

\begin{eqnarray}
& & \forall \beta \;\mbox{with}\;
0<\beta <C_0, \; \forall E \subset \Omega \times (0,T)\;
\mbox{with}\; |E|>0, \; \exists \epsilon\in \Upsilon\;
 \mbox{such that}\; \nonumber \\
& & 
\|w_\epsilon\|_{\infty,E}\le \beta
\label{eqc01}
\end{eqnarray}

\noindent Thus, taking $\beta=C_0-\frac{1}{n}$, $E=\Omega\times (0,T)$,
there exists $\epsilon_n$ a subsequence in $\Upsilon$ such that
$|w_{\epsilon_n}(y)| \le C_0-\frac{1}{n}, \; a.e. \; y\in E$.
This implies that $\|w_{\epsilon_n}\|_\infty \le C_0$ with $\epsilon_n
\in \Upsilon$. This contradicts (\ref{eqepsilon}). Therefore,
(\ref{eqc0}) is true.

We now prove that (\ref{eqgamma}) is true. Assume to the contrary that 
(\ref{eqgamma}) is not true. Then there is a subsequence
$\epsilon_k$ in $\Upsilon$ such that $m_{\epsilon_k}\rightarrow
_{k \rightarrow \infty} 0$. But we have

\begin{eqnarray*}
m_{\epsilon_k}&=&\int_{\Omega\times V\times (0,T)}
|\chi_{w_{\epsilon_k}(x,t)}(v)|dx dv dt =
\int_{\Omega\times (0,T)}
|w_{\epsilon_k}(x,t)|dx  dt 
\end{eqnarray*}

\noindent
Hence $\int_{\Omega\times (0,T)}
|w_{\epsilon_k}(x,t)|dx  dt \rightarrow 0$. Therefore there is a
subsequence $w_{\epsilon_{k_n}}$ that converges a.e. to $0$ on 
$\Omega\times (0,T)$. In particular, $w_{\epsilon_{k_n}}\rightarrow 0$
on $E$, where $E$ is the set given in (\ref{eqc0}).
Using Egoroff theorem \cite{evans}, $w_{\epsilon_{k_n}}\rightarrow 0$ almost
uniformly on $E$
(Recall from the remark after the statement (\ref{eqgamma}) that 
$E$ can be selected to satisfy $0<|E|<\infty$). 
That is, $\forall \eta>0, \; \exists E_\eta \subset E$
such that $|E\setminus E_\eta|<\eta$ and $w_{\epsilon_{k_n}}\rightarrow 0$
uniformly on $E_\eta$. Now fix $\eta>0$ small and let $\delta>0$ be
given, then there is $n^\prime$ depending on $\delta$ such that 

\begin{eqnarray}
& &
|w_{\epsilon_{k_n}}(y)|<\delta \;\;\; \forall y \in E_\eta, \;\; \forall
\epsilon_{k_n}<\epsilon_{k_{n^\prime}}
\label{equeps}
\end{eqnarray}

\noindent
Now let 

\begin{eqnarray}
& &
\tilde{E}=\{ x\in E: \; |w_{\epsilon_{k_n}}(x)|>\beta \;\;\;\; \forall  
 \epsilon_{k_n}<\epsilon_{k_{n^{\prime}}} \}
\label{eqtildee}
\end{eqnarray}

\noindent then (\ref{eqc0}) implies that $|\tilde{E}| >\alpha>0$ for some
$\alpha>0$. Now choose $\eta<\alpha$ then $E_\eta$ must contain a subset
$\hat {E}\subset \tilde{E}$ with $|\hat {E}|>0$. For otherwise 
the set $F=\tilde{E}\setminus \tilde{\tilde{E}}$ where

$$\tilde{\tilde{E}}=\{x\in \tilde{E}\cap E_\eta:
\; |w_{\epsilon_{k_n}}(x)|>\beta \; \forall  
 \epsilon_{k_n}<\epsilon_{k_{n^{\prime}}} \}, \;\;\;\; \mbox{and}\;\;
|\tilde{\tilde{E}}|=0$$

\noindent is included in $E\setminus
E_\eta$ ($F\subset E\setminus E_\eta$ ) and
$\alpha<|F|\le |E\setminus E_\eta|<\eta<\alpha$ which is impossible.
Now pick $\delta<\beta$ in  (\ref{equeps}). Then
in particular, we obtain 

$$\|w_{\epsilon_{k_n}}\|_{\infty,\hat{E}}<\beta \;\;\;\;
\forall  \epsilon_{k_n}<\epsilon_{k_{n^{\prime}}}$$

\noindent which is a contradiction to (\ref{eqtildee}). 
Therefore, (\ref{eqgamma}) is true.

Thus, we have $0<\gamma<m_\epsilon=|A_\epsilon| <C \;\; \forall 
\epsilon\in \Upsilon$ (Consult
(\ref{eqmepsilon}) and (\ref{eqgamma})). Now using 
the regularity of the Lebesgue measure, we have
for any $\eta>0$ such that $\gamma-\eta>0$, there exist
a compact set $F_\epsilon^\eta$ and an open set $\;U_\epsilon^\eta$
such that $F_\epsilon^\eta \subset A_\epsilon \subset U_\epsilon^\eta$
and  $|A_\epsilon|-\eta<|F_\epsilon^\eta|<|A_\epsilon|$ and
$|A_\epsilon|<|U_\epsilon^\eta|<|A_\epsilon|+\eta<C+\eta$. 
Thus for $\eta<\gamma/2$, we
can select $F_\epsilon^\eta$ and $\;U_\epsilon^\eta$ so that 

\begin{eqnarray}
& &
0<\gamma/2 <|F_\epsilon^\eta|\le |A_\epsilon| \le 
|U_\epsilon^\eta|<C+\gamma/2\; \;\; \forall \epsilon \in \Upsilon
\label{eqreg0}
\end{eqnarray}

Above we have used (\ref{eqmepsilon}) and (\ref{eqgamma}).
Now by Vitali's Covering Theorem \cite{evans}, there exists a countable
collection $G_\epsilon$ of disjoint closed balls in $U_\epsilon^\eta$ 
such that diam $B\le \eta$ for all $B\in G_\epsilon$ and 
$|U_\epsilon^\eta-\cup_{B\in G_\epsilon} B|=0$.
Using (\ref{eqreg0}) above, we then conclude that
$|\cup_{B\in G_\epsilon} B |$ is bounded below and above
by positive constants independent of $\epsilon \in \Upsilon$. 
Thus the projection $V_\epsilon$ of $A_\epsilon$ with respect to the $v$
axis has a one dimensional Lebesgue measure which is 
bounded above by a positive constant independent of $\epsilon \in
\Upsilon$. This proves the fact that
$n_\epsilon=|V_\epsilon|<C$ with $C$ a constant independent of $\epsilon \in
\Upsilon$.

Now we have

\begin{eqnarray}
& &
\int_{\Omega\times (0,T)}
\frac{1}{\epsilon}\int_0^t e^{(s-t)/\epsilon}
|\chi_{w_\epsilon(x,s)}|ds dt dx \nonumber \\
&=&
\int_{\Omega\times (0,T)} |\chi_{w_\epsilon(x,s)}|
(1-e^{(s-T)/\epsilon}) ds dx \nonumber
\end{eqnarray}

Thus, we have

\begin{eqnarray}
& &
\int_V (\int_{\Omega\times (0,T)}
\frac{1}{\epsilon}\int_0^t e^{(s-t)/\epsilon}
|\chi_{w_\epsilon(x,s)}|dx ds dt)^{1/p}dv \nonumber \\
&=&
\int_V (\int_{\Omega\times (0,T)} |\chi_{w_\epsilon(x,s)}|
(1-e^{(s-T)/\epsilon}) ds dx)^{1/p}dv \nonumber \\
&=&
\int_{V_\epsilon} (\int_{\Omega\times (0,T)} |\chi_{w_\epsilon(x,s)}|
(1-e^{(s-T)/\epsilon}) ds dx)^{1/p}dv \nonumber \\
&\le&
(\int_{V_\epsilon} \int_{\Omega\times (0,T)} |\chi_{w_\epsilon(x,s)}|
(1-e^{(s-T)/\epsilon}) ds dx dv)^{1/p} (n_\epsilon)^{1/p^\prime} \nonumber \\
&\le& C^{1/p} C^{1/p^\prime}=C
\end{eqnarray}

\noindent with $C$ independent of $\epsilon$. 
Above we have used Lemma \ref{lem2}, Holder inequality,
and the uniform boundedness of $n_\epsilon=|V_\epsilon|$.

Using this in (\ref{proot0}), we conclude that

\begin{eqnarray}
& &
\int_{V}
(\int_{\Omega\times (0,T)}
|g_\epsilon(x,v,t)|^p dx dt  )^{1/p}dv \nonumber \\
&\le & 4^{1/p}
\mbox{max} [ \int_V(\int_{\Gamma_0\times (0,T)}
|g_{\epsilon 0}(x,v,t)|^p)^{1/p}dv,
\int_V (\int_{\Omega}|g_{\epsilon}^0(x,v))|^pdx)^{1/p}dv,\nonumber \\
& & \int_V 
(\int_{\Gamma_1\times (0,T)}|g_{\epsilon 1}(x,v,t)|^pdx dv)^{1/p}dv,
C^{1/p} C^{1/p^\prime}]
\label{proot1}
\end{eqnarray}

On the other hand, using Minkowski inequality \cite{lieb}, we have

\begin{eqnarray}
& &
(\int_{\Omega \times (0,T)}
(\int_{V} |g_\epsilon|dv)^p dx dt )^{1/p} \nonumber \\
&\le & 
\int_{V}
(\int_{\Omega\times (0,T)}
|g_\epsilon(x,v,t)|^p dx dt  )^{1/p}dv 
\label{proot2}
\end{eqnarray}

Taking the limit as $p \rightarrow \infty$ in (\ref{proot1}) and
(\ref{proot2}), we conclude that 
$\|\int_{V} |g_\epsilon|dv\|_{L^\infty{(\Omega \times (0,T))}}$ is uniformly
in $\epsilon$ bounded and hence $w_\epsilon$ is also uniformly in
$\epsilon$ bounded in $L^\infty{(\Omega \times (0,T))}$. This concludes
the proof of (i).

\vskip .5cm
\noindent
{\it Second proof of the $L^\infty$ bound}
\vskip .3cm

Here, we shall assume that $|g_\epsilon^0(x,v)|\le 1$,
$|g_{\epsilon 0}(y,v,t)|\le 1$, $|g_{\epsilon 1}(y,v,t)|\le 1$. We shall
also assume as in \cite{bp1} 
that $g_\epsilon^0(x,v)\mbox{sign}(v)=|g_\epsilon^0(x,v)|$,
$g_{\epsilon 0}(y,v,t)\mbox{sign}(v)=|g_{\epsilon 0}(y,v,t)|$,
and $g_{\epsilon 1}(y,v,t)\mbox{sign}(v)=|g_{\epsilon 1}(y,v,t)|$.
Let $\tilde{v}$ denote a positive number such that
the support in $v$ of $g_\epsilon^0$, $g_{\epsilon 0}$, and $g_{\epsilon
1}$ is included in $[-\tilde{v}, \tilde{v}]$ (recall that we assumed
that these data have supports that are included in a fixed compact set
of $V$. Then using the sign condition on the data and the
integral representation we conclude that 
$g_\epsilon(x,v,t) \mbox{sign}(v)=|g_\epsilon(x,v,t)|$. Using
the fact that the data are bounded by $1$ and the integral
representation respectively in $\Omega_0$, $\Omega_{01}$, and
$\Omega_1$,
we obtain that $|g_\epsilon(x,v,t)| \le 1$. 

To obtain the uniform in $\epsilon$ bound of $w_\epsilon$, we use the
iterations (\ref{it1})-(\ref{it4}) and its corresponding 
integral representation

\begin{eqnarray*}
& & \mbox{In} \;\; \Omega_0\;\; \\ 
& &
\;\;g_\epsilon^1(x,v,t)=g_{\epsilon 0}(x_\star -\frac{x_1}{a_1(v)}a_\star(v),
v,t-\frac{x_1}{a_1(v)})\mbox{exp}({-x_1/(a_1(v)\epsilon)})
\\ & &
+\frac{1}{\epsilon} \int_{t-\frac{x_1}{a_1(v)}}^t
e^{(s-t)/\epsilon}\chi_{w_\epsilon^0(x(s),s)}(v)ds \\
& & \\
& & \mbox{In} \;\; \Omega_{01}\\
& &
\;\; g_\epsilon^1(x,v,t)=g^0_\epsilon(x-a(v)t,v)\mbox{exp}(-t/\epsilon)
+\frac{1}{\epsilon} \int_{0}^t
e^{(s-t)/\epsilon}\chi_{w_\epsilon^0(x(s),s)}(v)ds \\
& & \\
& & \mbox{In} \;\; \Omega_1 \;\; \\ 
& &
\;\; g_\epsilon^1(x,v,t)=g_{\epsilon 1}
(x_\star+\frac{1-x_1}{a_1(v)}a_\star(v),v,t-\frac{x_1-1}{a_1(v)})
\mbox{exp}{((1-x_1)/\epsilon a_1(v))} 
\\ & &
+\frac{1}{\epsilon} \int_{t-\frac{x_1-1}{a_1(v)}}^t
e^{(s-t)/\epsilon}\chi_{w_\epsilon^0(x(s),s)}(v)ds \\
\end{eqnarray*}

\noindent where $x(s)=x+(s-t)a(v)$, $x=(x_1,x_\star)$, and
$a(v)=(a_1(v),a_\star(v))$.

Let $w^0$ be an initial iterate such that $\|w^0\|_{L^\infty(\Omega
\times (0,T))}\le \tilde {v}$. Then by definition of $\tilde{v}$, we
have $g_{\epsilon 0}(y,v,t)=0$, $g_{\epsilon 1}(y,v,t)=0$,
$g_{\epsilon}^0(x,v)=0$, and $\chi_{w^0(x,t)}(v)=0$, for all 
$v$ with $|v|>\tilde{v}$. 

Now using the above integral representation,
we conclude that $g_\epsilon(x,v,t)=0$ for
$|v|>\tilde{v}$. Using this and the sign property of $g_\epsilon$
($|g_\epsilon(x,v,t)|=g_\epsilon(x,v,t)\mbox{sign}(v)$), we obtain

\begin{eqnarray*}
|w_\epsilon^1(x,t)|&=&|\int_V g_\epsilon(x,v,t)dv | \nonumber \\
&\le &\mbox{max}(|\int_{v>0} g_\epsilon(x,v,t)dv|, |\int_{v <0} |g_\epsilon
(x,v,t)|dv|) \nonumber \\
&\le & \tilde{v}
\end{eqnarray*}

Thus, the contraction operator maps elements $w^0$ with
$\|w^0\|_{L^\infty(\Omega \times (0,T))}<\tilde{v}$ into element with the same
property. Therefore the fixed point $w_\epsilon$ has also this property.
This concludes the proof of the uniform bound in $\epsilon$ of
$w_\epsilon$ in $L^\infty(\Omega \times (0,T))$.

Because of Lemma \ref{lem1} $g_\epsilon$ is uniformly
bounded in $L^\infty(\Omega \times V\times [0,T])$. Hence 
$w_\epsilon$ is uniformly bounded in $L^\infty(\Omega \times [0,T])$.

(ii) Now set
$w_\infty=\mbox{sup}_{\epsilon>0}
\|w_\epsilon\|_{L^\infty(\Omega \times [0,T])}$,
the terms $\chi_{w_\epsilon}$ in the integral
representation in Theorem \ref{teok} 
 are supported by $v \in [-w_\infty, w_\infty]$, the other terms are
supported by $v$ in the compact supports of the boundary and initial
data. Thus, for all $t\in [0,T]$, $g_\epsilon$ remains
compactly supported, with compact supports included in
$\mbox{Supp}_v g_\epsilon^0 \cup \mbox{Supp}_v g_{\epsilon 0}\cup 
\mbox{Supp}_v g_{\epsilon 1} \cup [-w_\infty,w_\infty]$,
which in turn are included in a fixed compact set independent of 
$\epsilon$. 

(iii)
Now set
$a_\infty=\mbox{sup}_{1\le i\le N, v\in S}|a_i(v)|$, with
$S=\mbox{Supp}_v g_\epsilon^0 \cup  \mbox{Supp}_v g_{\epsilon 1} 
\cup  \mbox{Supp}_v g_{\epsilon 0}\cup [-w_\infty,w_\infty]$. 
We conclude that
$\mbox{sup}_{1\le i\le N, v\in S^\prime}|a_i(v)|\le a_\infty$,
where
$S^\prime=\{v \in\mbox{supp}_v g_\epsilon(x,.,t),\; (x,t)\in \Omega \times
(0,T) \}$. And the lemma is proved.

\vskip 0.10in
In order to pass to the limit as the microscopic scale go to $0$,
we shall need to control the spatial and temporal variations of 
$g_\epsilon$ and $w_\epsilon$ in terms of $\epsilon$.
This is given in the following lemma.

\begin{lem}\label{lem4}
Assume that 

\begin{eqnarray}
& &
\|g_{\epsilon 0}\|_{L^\infty(\Gamma_0^{-}\times [0,T])}<C_1,\;\;
\|g^0_\epsilon\|_{L^\infty(\Omega\times V)}<C_2,\;\; 
\|g_{\epsilon 1}\|_{L^\infty(\Gamma_1^{-}\times [0,T])}<C_3,\nonumber \\
& &
\|g^0_\epsilon \|_{L^1(\Omega\times V)}<C_4,\;\; 
\|a(v)\cdot n g_{\epsilon 0}\|_{L^1(\Gamma_0^{-}\times (0,T))}<C_5,\;\; 
\|a(v)\cdot n g_{\epsilon 1}\|_{L^1(\Gamma_1^{-}\times (0,T))}<C_6  
\nonumber \\
& &
\|g_\epsilon^0\|_{L^1(V;BV_{loc}(\Omega))} <C_{7},\nonumber 
\end{eqnarray}

\noindent with $C_i,i=1,\cdots,7$ positive constants 
independent of $\epsilon$. 
Assume also that the initial and boundary data $f_{\epsilon
0}$, $g_\epsilon^0$, and $g_{\epsilon 1}$ are compactly supported in
$v\in V$ with
supports included in a fixed compact set independent of
$\epsilon$. 

\noindent Then 

1) $g_\epsilon(\cdot,\cdot,t)$ and $w_\epsilon(\cdot,t)$, 
$t\in [0,T]$ are uniformly bounded 
in $BV_{loc}(\Omega\times L^1(V))$
and $BV_{loc}(\Omega)$ respectively.

2)
$w_\epsilon$ is time Lipschitz continuous in $L^1_{loc}(\Omega)$
uniformly in $\epsilon$; i.e. for any open bounded subset $U$ of
$\Omega$ with $\bar{U}\subset \Omega$, we have

\begin{eqnarray}
& &
\|w_\epsilon(\cdot,t_2)-w_\epsilon(\cdot,t_1)\|_{L^1(U)}
<a_\infty\|g_\epsilon\|_{L^\infty([0,T];BV(U\times L^1(V)))} 
(t_2-t_1)<C(t_2-t_1),\nonumber \\
& & \forall \;0\le t_1<t_2\le T  \label{eqbvloct1}
\end{eqnarray}

\noindent where $C$ is a constant depending on $U$ but is
independent of $\epsilon$ and $a_\infty$ is introduced in the proof of
Lemma \ref{lemc1} above.

3) Under the additional assumption 

\begin{eqnarray}
& &
\|g_\epsilon^0(\cdot,\cdot)-\chi_{w^0(\cdot)}(\cdot)\|_
{L^1_{loc}(\Omega\times L^1(V))}
\rightarrow_{\epsilon \rightarrow 0} 0
\label{eqini0}
\end{eqnarray}

\noindent we can estimate the error 
between the kinetic solution and exact entropy solution as follows

\begin{eqnarray}
\|g_\epsilon -
\chi_{w_\epsilon}\|_{L^\infty([0,T];L^1_{loc}(\Omega\times L^1(V)))}
&\le& \epsilon a_\infty
\|g_\epsilon^0(x,v)\|_{BV_{loc}(\Omega\times L^1(V))}
\nonumber \\
& &
+\epsilon a_\infty
\|g_\epsilon(x,v,t)\|_{L^\infty([0,T];BV_{loc}(\Omega\times L^1(V))))}
\nonumber \\
& & +2 \|g_\epsilon^0(x,v)-\chi_{w^0(x)}\|_
{L^1_{loc}(\Omega\times L^1(V))}\nonumber \\
& &
    \rightarrow_{\epsilon \rightarrow 0} 0 
\label{eqbvloct5}
\end{eqnarray}

4) The function $w_\epsilon$
is uniformly bounded in $BV_{loc}(\Omega\times(0,T))$.

\end{lem}

\vskip .5cm
\noindent
{\bf Proof.}
\vskip .5cm

1) Let $0<t<T$ be fixed and $h>0$ be small. The case of $h<0$ will
be handled similarly.
Let $\tau_h^i g_\epsilon(x,v,t)=
g_\epsilon(x_1,\cdots,x_i+h e_i,\cdots,x_d,v,t)$, $i=1,\cdots,d$. 
Multiplying the equation (\ref{eq11}) for $\tau_h^1 g_\epsilon
-g_\epsilon$ by $\varphi$ 
with $\varphi$ a test function which is Lipschitz continuous
in $(0,1-h)\times \ir^{d-1}\times V\times [0,T]$ with compact support in
$x$ in $(0,1-h) \times \ir^{d-1}$ to be precised later, 
and integrating by parts, we obtain

\begin{eqnarray} 
& &
\int_{(0,1-h)\times \ir^{d-1}\times V}
((\tau_h^1 g_\epsilon-g_\epsilon)\varphi)
(\cdot,\cdot,t)-
\int_{(0,1-h)\times \ir^{d-1}\times V}
((\tau_h^1 g_\epsilon-g_\epsilon)\varphi)(\cdot,\cdot,0)
\nonumber \\
& &
-\int_{(0,1-h)\times \ir^{d-1}\times V\times (0,t)}
(\partial_t\varphi+a(v)\cdot \partial_x\varphi)
(\tau_h^1 g_\epsilon-g_\epsilon)
\nonumber \\
&=& \frac{1}{\epsilon}\int_{(0,1-h)\times \ir^{d-1}\times V\times (0,t)}
((\chi_{\tau_h^1 w_\epsilon}-\chi_{w_\epsilon})
-(\tau_h^1 g_\epsilon-g_\epsilon))\varphi
\label{ke1N}
\end{eqnarray}

We then take 
$\varphi=\mbox{sign}^{\mu} (\tau_h^1 g_\epsilon-g_\epsilon) \psi(x,t)$
with $x\mbox{sign}^{\mu} (x) \ge 0 \;\; x\in \ir$,
and $\psi$ is a nonnegative test function 
which is Lipschitz continuous in 
$(0,1-h)\times \ir^{d-1}\times V\times [0,T]$ with compact support in
$x$ in $(0,1-h) \times \ir^{d-1}$
and $\mbox{sign}^{\mu}$ is a regularization of $\mbox{sign}$
function. Proceeding as in the proof of Theorem \ref{teok}, we obtain

\begin{eqnarray*}
& &
\int_{(0,1-h)\times \ir^{d-1}\times V}|\tau_h^1 g_\epsilon-g_\epsilon| 
\psi(\cdot,\cdot,t)
-\int_{(0,1-h)\times \ir^{d-1}\times V}|\tau_h^1 g_\epsilon-g_\epsilon| 
\psi (\cdot,\cdot,0)
\nonumber \\
& &
-\int_{(0,1-h)\times \ir^{d-1}\times V\times (0,t)}
(\partial_t \psi+a(v)\cdot \partial_x \psi)
|\tau_h^1 g_\epsilon-g_\epsilon|
\nonumber \\
&=& \frac{1}{\epsilon}\int_{(0,1-h)\times \ir^{d-1}\times V\times (0,t)}
((\chi_{\tau_h^1 w_\epsilon}-\chi_{w_\epsilon})-
(\tau_h^1 g_\epsilon-g_\epsilon))
\mbox{sign}(\tau_h^1 g_\epsilon-g_\epsilon)\psi \nonumber \\
&\le& 0
\end{eqnarray*}

\noindent where in the last inequality we have used 
the properties of $\chi$, 
we then have

\begin{eqnarray*}
& &
\int_{(0,1-h)\times \ir^{d-1}\times V} \psi|\tau_h^1 g_\epsilon-g_\epsilon| 
(\cdot,\cdot,t)
\nonumber \\
&\le & 
\int_{(0,1-h)\times \ir^{d-1}\times V}\psi |\tau_h^1 g_\epsilon-g_\epsilon| 
(\cdot,\cdot,0)+
\int_{(0,1-h)\times \ir^{d-1}\times V\times (0,t)}
(\partial_t \psi+a(v)\cdot \partial_x \psi)
|\tau_h^1 g_\epsilon-g_\epsilon|
\end{eqnarray*}

\noindent In particular we have

\begin{eqnarray}
& &
\int_{O \times V}\psi |\tau_h^1 g_\epsilon-g_\epsilon| 
(\cdot,\cdot,t)
\nonumber \\
&\le & 
\int_{O\times V}\psi|\tau_h^1 g_\epsilon-g_\epsilon| 
(\cdot,\cdot,0)+
\int_{O \times V\times (0,t)}
(\partial_t \psi+a(v)\cdot \partial_x \psi)
|\tau_h^1 g_\epsilon-g_\epsilon|
\label{eqbvloc0}
\end{eqnarray}

\noindent
for any open set with $\bar{O}\subset (0,1-h)\times \ir^{d-1}$
and $\psi$ any Lipschitz continuous function 
in $O \times V\times [0,T]$ with compact support in
$x$ in $O$. Similarly, we
have for $i=2,\cdots, d$

\begin{eqnarray}
& &
\int_{O \times V}\psi |\tau_h^i g_\epsilon-g_\epsilon| 
(\cdot,\cdot,t)
\nonumber \\
&\le & 
\int_{O\times V}\psi|\tau_h^i g_\epsilon-g_\epsilon| 
(\cdot,\cdot,0)+
\int_{O \times V\times (0,t)}
(\partial_t \psi+a(v)\cdot \partial_x \psi)
|\tau_h^i g_\epsilon-g_\epsilon|
\label{eqbvloc0n}
\end{eqnarray}

\noindent
for any open set with $\bar{O}\subset (0,1)\times \ir^{d-1}$
and $\psi$ any Lipschitz continuous function 
in $O \times V\times [0,T]$ with compact support in
$x$ in $O$.

Let $i\in \{2,\cdots,d\}$ be fixed.
Let $U$ and $O$ be open bounded subsets of $\Omega$ such that
$\bar{U}\subset O \subset \bar{O} \subset \Omega$.
Let $\psi$ be a Lipschitz continuous function
in $O \times V\times [0,T]$ with compact support in
$x$ in $O$
such that $U\subset \mbox{supp}_x \psi
\subset O$. Then (\ref{eqbvloc0n}) holds for such $\psi$ and $O$.

We wish to prove that 

\begin{equation}
\int_{U\times V} |\tau^i_h g_\epsilon-g_\epsilon| \le Ch
\label{eqbvg1}
\end{equation}

\noindent where $C$ depends on $U$ but is independent of
$\epsilon$. It is enough to prove this relation for $U$ of the form
$U=(y_1-\alpha,y_1+\alpha)\times B(y_\star,R)$ where $\alpha>0$ and
$y=(y_1,y_\star)\in \Omega$ are arbitrary elements of $\ir^{+\star}$
and $\Omega$ such that $0<y_1-\alpha<y_1+\alpha<1$ and 
$R>0$ is arbitrary radius.
Let $\beta>0$ and $\gamma>0$ be such that 
$0<y_1-\alpha-\beta-\gamma<y_1+\alpha+\beta+\gamma<1$.
Let $0<t_1<T$ be such that $a_\infty t_1=\beta$.
Let $O=(y_1-\alpha-a_\infty t_1-\gamma,y_1+\alpha +a_\infty t_1+\gamma)\times 
B(y_\star,R+\delta+d a_\infty t_1)$, with $\delta>0$.
Let $t\in (0,t_1]$.
Consider now the functions

\[
\varphi_1(x_1,\tau)=\left \{ \begin{array}{lllll}
0 & 0\le x_1<y_1-\alpha-a_\infty(t-\tau)-\gamma \\
  & 0\le \tau \le t \\
\frac{1}{\gamma}(x_1-y_1+\alpha+a_\infty(t-\tau))+1 
& y_1-\alpha-a_\infty(t-\tau)-\gamma \le
x_1<   \\
  & y_1-\alpha-a_\infty(t-\tau), \;\;\; 0\le \tau \le t \\
1 & y_1-\alpha-a_\infty(t-\tau)\le x_1<y_1+\alpha+a_\infty(t-\tau)\\
  & 0\le \tau \le t \\
\frac{1}{\gamma}(y_1+\alpha+a_\infty(t-\tau)-x_1)+1 
& y_1+\alpha+a_\infty(t-\tau)\le x_1<  \\
  & y_1+\alpha+a_\infty(t-\tau)+\gamma, \;\;\; 0\le \tau \le t \\
0
 &y_1+\alpha+a_\infty(t-\tau) +\gamma\le x_1 \le 1 \\
  & 0\le \tau \le t 
\end{array}
\right. 
\]

\noindent and

\[
\varphi_2(x_\star,\tau)=\left \{ \begin{array}{lllll}
1 & 0\le |x_\star-y_\star| <R+d a_\infty (t-\tau) \\
  & 0\le \tau \le t \\
\frac{1}{\delta}(R+da_\infty(t-\tau)-|x_\star-y_\star|)+1
& R+da_\infty (t-\tau)\le |x_\star-y_\star|< \\
 & R+da_\infty(t-\tau)+\delta, \;\;\; 0\le \tau \le t \\
0 & R+da_\infty(t-\tau)+\delta \le |x_\star-y_\star| \\
  & 0\le \tau \le t 
\end{array}
\right. 
\]

Now let $\psi(x,\tau)=\varphi_1(x_1,\tau)\varphi_2(x_\star,\tau)$, 
$\tau \in [0,t]$ and $x=(x_1,x_\star)$. It is clear that $\psi$ is
nonnegative Lipschitz continuous function in
$O\times V\times [0,t]$ with compact support in $x$ in $O$ 
and $U\subset \mbox{supp}_x \psi \subset O$.
Thus, plugging $\psi$ in (\ref{eqbvloc0n}) and using the fact that
$g_\epsilon^0$ is uniformly bounded in 
$BV_{loc}(\Omega\times L^1(V))$ (since $g_\epsilon^0$ is
uniformly bounded in 
$L^1(V;BV_{loc}(\Omega)) \subset BV_{loc}(\Omega\times L^1(V))$)
yields (\ref{eqbvg1})
for $t\in (0,t_1]$. Now let $t_2>t_1$ be such that
$a_\infty(t_2-t_1)=\beta$. Proceeding as above and using the fact that 
$g_\epsilon(\cdot,\cdot,t_1)$ is uniformly bounded in
$BV_{loc}(\Omega\times L^1(V))$, we conclude that
$g_\epsilon(\cdot,\cdot,t)$ is uniformly bounded in 
$BV_{loc}(\Omega\times L^1(V))$ for any $t\in (t_1,t_2]$.
Continuing this process we conclude that 
$g_\epsilon(\cdot,\cdot,t)$ is uniformly bounded in 
$BV_{loc}(\Omega\times L^1(V))$ for any $t\in [0,T]$.

Finally, using similar constructions we can prove that for any open
bounded subset $O$ of $(0,1-h)\times \ir^{d-1}$ with
$\bar{O} \subset (0,1-h)\times \ir^{d-1}$, we have 

\[
\int_{O\times V} |\tau_h^1 g_\epsilon -g_\epsilon|\le Ch
\]

\noindent where $C$ is a positive constant depending on $O$, but 
is independent of $\epsilon$. This concludes the proof that
$g_\epsilon$ is uniformly bounded in
$L^\infty([0,T];BV_{loc}(\Omega\times L^1(V)))$.
The uniform bound of $w_\epsilon$ in 
$L^\infty([0,T];BV_{loc}(\Omega))$ can then be deduced from 
that of $g_\epsilon$.
And the statement 1) is proved.

2)
Let $0\le t_1<t_2\le T$ and $U$ be an open bounded subset of $\Omega$
with $\bar{U} \subset \Omega$. 
Let $\psi(x) \in C_0^1(U)$. Multiplying Eq. (\ref{eq11}) by $\psi$
and integrating over $U \times (t_1,t_2)\times V$, we obtain

\[
\int_{U \times V\times (t_1,t_2)}\partial_t g_\epsilon \psi+
\sum_i \int_{U \times V\times (t_1,t_2)} a_i(v) \partial_{x_i}
g_\epsilon \psi=\frac{1}{\epsilon}\int_{U \times V\times (t_1,t_2)}
(\chi_{w_\epsilon}-g_\epsilon)\psi=0 
\]

\noindent Hence, we have

\begin{eqnarray}
& & \int_{U}(w_\epsilon(x,t_2)-w_\epsilon(x,t_1))\psi(x)=
-\int_{t_1}^{t_2}\sum_i \int_{U \times V} a_i(v) \partial_{x_i}
g_\epsilon \psi \label{eqbvloct0}
\end{eqnarray}

Since $\partial_{x_i}g_\epsilon$,
$i=1,\cdots,d$ are locally finite measures (consult 1) above),
the integrand on the right side is bounded by $a_\infty C(U)$
for $|\psi (x)|\le 1$.
Taking the supremum of (\ref{eqbvloct0}) over all $\psi$ with
$|\psi(x)|\le 1$ yields (\ref{eqbvloct1}).

3)  Let $U$ be an open bounded set of $\Omega$ such that
$\bar{U}\subset \Omega$. Let $O$ be an open bounded 
set of $\Omega$ such that 
$\bar{U} \subset O \subset \bar{O} \subset \Omega$. 
Taking $G_\epsilon(x,v,t)=g_\epsilon(x,v,t+\Delta t)$ and proceeding as in the
derivation of (\ref{estcont}) and the proof of the uniform $BV_{loc}$
bound (consult part 1) above), we obtain

\begin{equation}
\int_{U\times V}
|g_\epsilon(x,v,t+\Delta t)-g_\epsilon(x,v,t)|
\le  \int_{O\times V}
|g_\epsilon(x,v,\Delta t)-g_\epsilon(x,v,0)|
\end{equation}

\noindent from which we deduce

\begin{equation}
\|\partial_t g_\epsilon(x,v,t)\|_{L^1(U\times V)}
\le \|\partial_t g_\epsilon(x,v,t=0)\|_{L^1(O\times V)}
\label{eqlip0}
\end{equation}

The kinetic equation (\ref{eq11}) yields

\begin{eqnarray}
& & \|\partial_t g_\epsilon(x,v,t=0)\|_{L^1(O\times V)} \nonumber \\
&\le& \|(a(v)\cdot \partial_x) g_\epsilon(x,v,t=0)\|_{L^1(O\times V)}+
\frac{1}{\epsilon} \|\chi_{w_\epsilon(x,t=0)}-g_\epsilon(x,v,t=0)\|_
{L^1(O\times V)} \nonumber \\
&\le& a_\infty \|g_\epsilon(x,v,t=0)\|_{BV(O\times L^1(V))}
+\frac{2}{\epsilon} \|g_\epsilon^0(x,v)-\chi_{w^0(x)}\|_
{L^1(O\times V)} \label{eqlip1}
\end{eqnarray}

Using again the kinetic equation (\ref{eq11}) together with (\ref{eqlip0}),
(\ref{eqlip1})  and the uniform
bound of $g_\epsilon(x,v,t)$ in
$L^\infty([0,T],BV_{loc}(\Omega\times L^1(V)))$,
we obtain

\begin{eqnarray}
& &\|g_\epsilon(x,v,t)-\chi_{w_\epsilon(x,t)}(v)\|_{L^1(U\times V)}
\nonumber \\
&\le&\epsilon\|\partial_t g_\epsilon(x,v,t)\|_{L^1(U\times V)}+
\epsilon \|(a(v)\cdot \partial_x) g_\epsilon(x,v,t)\|_{L^1(U\times V)}
\nonumber\\
&\le& \epsilon a_\infty \|g_\epsilon(x,v,t=0)\|_{BV(O\times L^1(V))}
+\epsilon a_\infty \|g_\epsilon(x,v,t)\|_{BV(U\times L^1(V))} \nonumber \\
& & +2 \|g_\epsilon^0(x,v)-\chi_{w^0(x)}\|_{L^1(O\times V)} \label{eqlip2}
\end{eqnarray}

Now, (\ref{eqlip2}) and (\ref{eqini0}) yield as $\epsilon \rightarrow 0$

\[
\|g_\epsilon(x,v,t)-\chi_{w_\epsilon(x,t)}(v)\|_{L^1(U\times V)}
\rightarrow 0
\]

The proof of 3) is now complete.

4) The proof is an immediate consequence of a combination of
1) and 2) above.

\begin{rema}\label{control0}
Notice that Lemma \ref{lem4} part 1) furnishes a local uniform in $\epsilon$
bound on the spatial variation on the microscopic scale. However, the 
local Lipschitz continuity is obtained only at the macroscopic level;
consult Lemma \ref{lem4} part 2). The temporal variation at the
microscopic level cannot, in general, be bounded uniformly in
$\epsilon$. Such uniform control can be achieved only if we can prevent
the possibility of a kinetic layer in (\ref{eq11}) (Consult Theorem
\ref{teokcln} and the remark before it).
\end{rema}

\section{Hydrodynamic limit of the kinetic problem and existence theory
for the conservation laws}

In this section we shall prove that the conservation laws
(\ref{eq3n})-(\ref{eq6}) has a solution in the sense of Definition 
\ref{defi1} below which selects a 
physically correct solution to this problem.

\begin{defi} \label{defi1}
We say that $w\in BV_{loc}(\Omega \times(0,T)) \cap 
L^\infty(\Omega \times [0,T])$
is a weak entropic solution of the problem
(\ref{eq3n})-(\ref{eq6}) if we have

\begin{eqnarray}
& & -\int_{\Omega\times (0,T)}(|w-k| \partial_t
\psi+\mbox{sign}(w-k)(A(w)-A(k))\cdot \nabla_x \psi) \nonumber \\
& &
+\int_{\Gamma_1 \times (0,T)}\psi
\mbox{sign}(w_1-k)((A(w_1)\cdot n)^- - (A(k)\cdot n)^-) \nonumber \\
& & 
+\int_{\Gamma_0^- \times(0,T)} a(v)\cdot n \psi |g_0-\chi_k| 
\le 0 \nonumber \\
& & \nonumber \\
& & \forall \psi \in
C^1_0(\bar{\Omega}\times V \times (0,T)),\;
\psi \ge 0,\;  \forall k \in \ir \nonumber 
\end{eqnarray}

and $w$ satisfies the initial condition

\[
w(x,0)=w^0(x)\;\;   \mbox{in}\;\; \Omega
\]

\end{defi}

We now state the following theorem about the existence of a solution to
the conservation laws.

\begin{teo}\label{teokcl}
Assume that 

\begin{eqnarray}
& &
\|g_{\epsilon 0}\|_{L^\infty(\Gamma_0^{-}\times [0,T])}<C_1,\;\;
\|g^0_\epsilon\|_{L^\infty(\Omega\times V)}<C_2,\;\; 
\|g_{\epsilon 1}\|_{L^\infty(\Gamma_1^{-}\times [0,T])}<C_3,\nonumber \\
& &
\|g^0_\epsilon \|_{L^1(\Omega\times V)}<C_4, \;\;
\|a(v)\cdot n g_{\epsilon 0}\|_{L^1(\Gamma_0^{-}\times (0,T))}<C_5,\;\; 
\|a(v)\cdot n g_{\epsilon 1}\|_{L^1(\Gamma_1^{-}\times (0,T))}<C_6  
\nonumber \\
& &
\|g_\epsilon^0\|_{L^1(V;BV_{loc}(\Omega))} <C_{7} \nonumber
\end{eqnarray}

\noindent with $C_i,i=1,\cdots,7$ positive constants 
independent of $\epsilon$. 

Assume also that the initial and boundary data $f_{\epsilon
0}$, $g_\epsilon^0$, and $g_{\epsilon 1}$ are compactly supported in
$v\in V$ with
supports included in a fixed compact set independent of
$\epsilon$. Finally assume that as $\epsilon \rightarrow 0$, 

\begin{eqnarray}
& &
\|w_\epsilon(\cdot,0)-w^0(\cdot)\|_
{L^1_{loc}(\Omega)}
=\|\int_V g_\epsilon^0(\cdot,v)-w^0(\cdot)\|_
{L^1_{loc}(\Omega)}
\rightarrow 0
\label{eqini1} \\
& & a(v)\cdot n g_{\epsilon 0} \rightarrow a(v)\cdot n g_{0}
\;\;\mbox{strongly in}\;\; L^1(\Gamma_0^- \times (0,T))
\label{dat3} \\
& & a(v)\cdot n g_{\epsilon 1} \rightarrow a(v)\cdot n g_{1}=a(v)\cdot
n\chi_{w_1} 
\;\;\mbox{strongly in}\;\; L^1(\Gamma_1^- \times (0,T))
\label{dat4} 
\end{eqnarray}

\noindent
Then $w_\epsilon$ converges strongly in $L^1(\Omega \times (0,T))$,
as $\epsilon$ goes to $0$, to an entropic solution of the problem
(\ref{eq3n})-(\ref{eq6}) in the sense of Definition \ref{defi1}.
\end{teo}

Before we give the proof of Theorem \ref{teokcl}, we shall state and
prove a preliminary result showing compactness of 
$w_\epsilon$ and $g_\epsilon$ respectively in $L^1(\Omega \times (0,T))$
and $L^1(\Omega \times V\times (0,T))$. 
We shall assume that $\Omega=(0,1)$.
It is not difficult to generalize our proof to the
case $\Omega=(0,1)\times \ir^{d-1}$.

\begin{lem}\label{lem7}
Assume that all assumptions of Theorem \ref{teokcl} hold.
Then 

\noindent i)
A subsequence of $w_\epsilon$ (still denoted $w_\epsilon$) converges
as $\epsilon \rightarrow 0$ to $w$ in 
$L^1_{loc}(\Omega \times (0,T))\cap
L^\infty([0,T];L^1_{loc}(\Omega))$
and in $L^\infty(\Omega\times [0,T])$ weak-$\star$.
Moreover $w_\epsilon$ converges
a.e. to $w$ in $\Omega \times(0,T)$ and
$w\in BV_{loc}(\Omega \times (0,T))$.

\noindent ii)
The $L^1_{loc}$ convergence of $w_\epsilon$ takes place actually in 
$L^1(\Omega \times(0,T))\cap L^\infty([0,T];L^1(\Omega))$.

\noindent iii)
Finally, we have $\|g_\epsilon -\chi_w\|_{L^1(\Omega \times V \times (0,T))} 
\rightarrow 0$ as $\epsilon \rightarrow 0$.

\end{lem}

To prove Lemma \ref{lem7} part ii), we shall also need the following
result.
 
\begin{teo}\label{lemglo}
Let $U$ be a bounded open subset of $\ir^N$ and let $v_n$ be a sequence
in $L^1_{loc}(U)$. Assume that as $n
\rightarrow \infty$, the sequence $v_n$ converges strongly in
$L^1_{loc}(U)$ to $v\in L^1_{loc}(U)$.
If $v_n$ is uniformly bounded in $L^\infty(U)$ then 
$v_n$ converges strongly to $v$ in $L^1(U)$.
\end{teo}

\vskip .25cm 
\noindent {\bf Proof of Theorem \ref{lemglo}}
\vskip .25cm 

Let $\eta>0$ be fixed. Since $U$ is bounded
there exists  a compact set $K_\eta \subset U$ such that the Lebesgue
measure $\mbox{meas}(U\setminus K_\eta) <\eta$. On the other hand
since $v_n$ is uniformly bounded in $L^\infty(U)$, by diagonal process
to pass to a further subsequence if necessary and uniqueness of the
limit, $v_n$ converges in $L^\infty$ weak-$\star$ to $v$. Hence
$v\in L^\infty(U)$. Now

\begin{eqnarray*}
\int_{U} |v_n-v| &=& \int_{U\setminus K_\eta} |v_n-v|+
\int_{K_\eta} |v_n-v| \\
&\le& \|v_n -v\|_\infty \mbox{meas}(U\setminus K_\eta)+
\int_{K_\eta} |v_n-v|  \\
&\le& C \eta+
\int_{K_\eta} |v_n-v|  \\
\end{eqnarray*}

\noindent where $C$ is a constant independent of $n$ and $\eta$.
Therefore since  
$\mbox{lim}_{n \rightarrow \infty}\int_{K_\eta} |v_n-v|=0$,

\begin{eqnarray*}
\mbox{lim sup}_{n \rightarrow \infty}\int_{U} |v_n-v|
&\le& C \eta
\end{eqnarray*}

This proves the statement since $\eta$ is arbitrary.

\vskip 1.0cm
{\bf Proof of Lemma \ref{lem7}}
\vskip 1.0cm

Using Lemma \ref{lem4} part 4) and Lemma \ref{lem2}
$w_\epsilon$ is bounded uniformly in
$L^1 \cap BV_{loc}(\Omega \times (0,T))$.
Hence
a subsequence of $w_\epsilon$ (still denoted 
$w_\epsilon$) converges to $w$ in $L^1_{loc}(\Omega \times (0,T))$ and 
almost everywhere in $\Omega \times (0,T)$.  
Moreover $w\in BV_{loc}(\Omega\times (0,T))$.
Using Lemma \ref{lemc1} and diagonal process
to extract a further subsequence, if necessary, $w_\epsilon$ 
converges in $L^\infty(\Omega \times [0,T])$ weak-$\star$ to a function 
$w \in L^\infty(\Omega \times [0,T])$.
Since $\Omega \times (0,T)$
is bounded the limit $w$ is in $L^1(\Omega \times (0,T))$. 
Now by the dominated convergence theorem
and the above, the convergence of $w_\epsilon$
takes place in fact in $L^1(\Omega \times (0,T))$. 

Now by Lemma \ref{lem4} part 1) $w_\epsilon(\cdot,t), \; t\in [0,T]$
is uniformly bounded in $BV_{loc}(\Omega)$. Hence it is precompact in
$L^1_{loc}(\Omega)$. Using Lemma \ref{lem4} part 2), 
$\|w_\epsilon(x,t)\|_{L^1(\Omega)}$ is Lipschitz continuous in time. By
diagonal process to extract a further subsequence, if necessary,
$w_\epsilon \rightarrow_{\epsilon \rightarrow 0} w$ strongly in 
$L^\infty([0,T];L^1_{loc}(\Omega))$. Now by the same process we used to
prove the strong $L^1$ convergence of $w_\epsilon$ to $w$ in
$L^1(\Omega \times (0,T))$, we conclude that 
$w_\epsilon \rightarrow_{\epsilon \rightarrow 0} w$ strongly in 
$L^\infty([0,T];L^1(\Omega))$.

By the properties of $\chi$, we conclude that $\chi_{w_\epsilon}$
strongly converges to $\chi_w$ in $L^1$. Using this and the integral
representation (Theorem \ref{teok}), and recalling that the boundary
data satisfy (\ref{dat3})-(\ref{dat4}), we infer that $g_\epsilon$ 
strongly converges to $\chi_w$ in $L^1$.
This concludes the proof of the lemma.

\vskip .1in

\vskip .5cm
\noindent
{\bf Proof of Theorem \ref{teokcl}}
\vskip .5cm

Using Lemma \ref{lem7} 
a subsequence of $w_\epsilon$ 
(still denoted $w_\epsilon $) converges strongly in
$L^1$ to $w$.
We know that $w\in BV_{loc}(\Omega \times(0,T)) \cap L^\infty(\Omega
\times [0,T])$ (consult the proof of Lemma \ref{lem7}).
Using Theorem \ref{teoent},  we have

\begin{eqnarray}
& &
-\int_{\Omega \times V\times (0,T)}
(\partial_t+a(v)\cdot \partial_x)(\psi)|g_\epsilon-\chi_k|
+\int_{\Gamma_0^-\times (0,T)} a(v)\cdot n 
\psi|g_{\epsilon 0}-\chi_k| \nonumber \\
& &
+\int_{\Gamma_1^{-}\times (0,T)} a(v)\cdot n 
\psi|g_{\epsilon 1}-\chi_k|
\le 0 \;\;\;\; \forall \psi \in C_0^1(\bar{\Omega} \times
(0,T)),\; \psi \ge 0,\;\forall k\in \ir \nonumber 
\end{eqnarray}

Using Lemma \ref{lem7}, Lemma \ref{lemc1}, (\ref{dat3}) and
(\ref{dat4}), and the properties of $\chi$, we then obtain

\begin{eqnarray}
& &
-\int_{\Omega\times (0,T)}
\partial_t \psi|w-k|
-\int_{\Omega\times (0,T)}
\mbox{sign}(w-k) (A(w)-A(k))\cdot \partial_x \psi 
\nonumber \\
& &
+\int_{\Gamma_0^-\times (0,T)} a(v)\cdot n 
\psi|g_0-\chi_k| 
+
\int_{\Gamma_1 \times (0,T)}
\mbox{sign}(w_1-k)((A(w_1)\cdot n)^- - (A(k)\cdot n)^-)\psi \nonumber \\
&\le &0  \;\;\;\;\;\;\;\;\;\;\;\;\;\;\;\;\;\;
\forall \psi \in C_0^1(\bar{\Omega}_g \times
(0,T)),\; \psi \ge 0,\;\forall k\in \ir \label{eqcons}
\end{eqnarray}

Finally, thanks to Lemma \ref{lem7}, (\ref{estcont}), and (\ref{eqini1}),
$w$
satisfies the initial conditions (\ref{eq6}).
Thus, combining (\ref{eqcons}) and the above, 
it is clear that $w$ is an entropic
solution in the sense of Definition \ref{defi1}
to the problem (\ref{eq3n})-(\ref{eq6}).

The proof of the theorem is now complete.

\vskip .5cm
As we saw in Remark \ref{control0},
the temporal variation at the
microscopic level cannot, in general, be bounded uniformly in
$\epsilon$. Such uniform control can be achieved only if we can prevent
the possibility of a kinetic layer in (\ref{eq11}). For this purpose, we
shall prepare the kinetic initial data so that $\frac{\partial}{\partial t}
g_\epsilon$ is uniformly bounded in $\epsilon$ and $t$, in particular at
$t=0$. In such case no kinetic initial layer will be present.
We therefore assume that the kinetic initial data satisfies \cite{bp}

\begin{eqnarray*}
\|g_\epsilon^0(\cdot,\cdot)-\chi_{w^0(\cdot)}(\cdot)\|_
{L^1_{loc}(\Omega\times L^1(V))}
\rightarrow_{\epsilon \rightarrow 0} 0
\end{eqnarray*}

\begin{teo}\label{teokcln}
Assume that 

\begin{eqnarray}
& &
\|g_{\epsilon 0}\|_{L^\infty(\Gamma_0^{-}\times [0,T])}<C_1,\;\;
\|g^0_\epsilon\|_{L^\infty(\Omega\times V)}<C_2,\;\; 
\|g_{\epsilon 1}\|_{L^\infty(\Gamma_1^{-}\times [0,T])}<C_3,\nonumber \\
& &
\|g^0_\epsilon \|_{L^1(\Omega\times V)}<C_4, \;\;
\|a(v)\cdot n g_{\epsilon 0}\|_{L^1(\Gamma_0^{-}\times (0,T))}<C_5,\;\; 
\|a(v)\cdot n g_{\epsilon 1}\|_{L^1(\Gamma_1^{-}\times (0,T))}<C_6  
\nonumber \\
& &
\|g_\epsilon^0\|_{L^1(V;BV_{loc}(\Omega))} <C_{7} \nonumber
\end{eqnarray}

\noindent with $C_i,i=1,\cdots,7$ positive constants 
independent of $\epsilon$. 

Assume also that the initial and boundary data $f_{\epsilon
0}$, $g_\epsilon^0$, and $g_{\epsilon 1}$ are compactly supported in
$v\in V$ with
supports included in a fixed compact set independent of
$\epsilon$. Finally assume that as $\epsilon \rightarrow 0$, 

\begin{eqnarray}
& &
\|g_\epsilon^0(\cdot,\cdot)-\chi_{w^0(\cdot)}(\cdot)\|_
{L^1_{loc}(\Omega\times L^1(V))}
\rightarrow_{\epsilon \rightarrow 0} 0
\label{eqini2} \\
& & a(v)\cdot n g_{\epsilon 0} \rightarrow a(v)\cdot n g_{0}
\;\;\mbox{strongly in}\;\; L^1(\Gamma_0^- \times (0,T))
\label{dat3n} \\
& & a(v)\cdot n g_{\epsilon 1} \rightarrow a(v)\cdot n g_{1}=a(v)\cdot
n\chi_{w_1} 
\;\;\mbox{strongly in}\;\; L^1(\Gamma_1^- \times (0,T))
\label{dat4n} 
\end{eqnarray}

\noindent
Then $g_\epsilon$ converges strongly in $L^\infty([0,T];L^1(\Omega
\times V))$, as $\epsilon$ goes to $0$, to $\chi_w$ and
$w$ is an entropic solution of the problem
(\ref{eq3n})-(\ref{eq6}) in the sense of Definition \ref{defi1}.
\end{teo}

Before we give the proof of Theorem \ref{teokcln}, we shall state and
prove the lemma below. 

\begin{lem}\label{lem7n}
Assume that all assumptions of Theorem \ref{teokcln} hold.
Then i) and ii) of Lemma \ref{lem7} hold true.
Moreover, we have $\|g_\epsilon -\chi_w\|_{L^\infty([0,T];L^1(\Omega
\times V))} \rightarrow 0$ as $\epsilon \rightarrow 0$.
\end{lem}

\vskip 1.0cm
{\bf Proof of Lemma \ref{lem7n}}
\vskip 1.0cm

We only need to prove the last statement in the lemma.
By Lemma \ref{lem4} part 3) 

\[
\|g_\epsilon -\chi_{w_\epsilon}\|_
{L^\infty([0,T];L^1_{loc}(\Omega\times L^1(V)))}
    \rightarrow_{\epsilon \rightarrow 0} 0 
\]

Thus 

\[
\|g_\epsilon -\chi_{w}\|_
{L^\infty([0,T];L^1_{loc}(\Omega\times L^1(V)))}
    \rightarrow_{\epsilon \rightarrow 0} 0 
\]

Since $g_\epsilon$ is uniformly bounded in $L^\infty(\Omega \times V
\times [0,T])$ (Lemma \ref{lem1}) 
and remains compactly supported in $v$ with support
included in a fixed compact set independent of $\epsilon$ 
(Lemma \ref{lemc1}), and
$g_\epsilon$ converges to $g$ in 
$L^\infty([0,T];L^1_{loc}(\Omega\times L^1(V)))$, we can apply Theorem
\ref{lemglo} to infer that 
$g_\epsilon \rightarrow \chi_w$ in 
$L^\infty([0,T];L^1(\Omega\times L^1(V)))$.
This concludes the proof of the lemma.

\vskip .1in

\vskip .5cm
\noindent
{\bf Proof of Theorem \ref{teokcln}}
\vskip .5cm

The proof of this theorem is similar to that of Theorem 
\ref{teokcl} and will not be repeated. 
$time$. 

\begin{rema}\label{rema2} Theorems \ref{teokcl} and \ref{teokcln}
are obtained under 
various assumptions including the assumptions that
the data $g_{\epsilon 0},g^0_\epsilon,$ and $g_{\epsilon 1}$ 
are compactly supported in $v$. In fact these theorems are
also valid when these data are not
necessarily compactly supported in $v$. The proof is based on 
a BV-regularization argument.
\end{rema}

\section{Cancellation of microscopic oscillations via the compensated
compactness} 

In this section we study the one-dimensional scalar
conservation law

\begin{eqnarray}
& & \partial_t w+\partial_x A(w) =0 
\;\; \mbox{in}\;\; \Omega\times (0,T) \label{eq3ncc}\\
& & \mbox{Boundary conditions for $w$ on $\Gamma_0\times (0,T)$
and $\Gamma_1\times (0,T)$}
\label{eq4cc}\\
& & \nonumber \\
& & w(x,0)=w^0(x) \;\;\mbox{in}\;\; \Omega \label{eq6cc}
\end{eqnarray}

The corresponding kinetic equation \cite{bp} is

\begin{eqnarray}
& & [\partial_t+a(v)\cdot \partial_x]
g_{\epsilon}(x,v,t)=\frac{1}{\epsilon}
(\chi_{w_{\epsilon}(x,t)}(v)-g_{\epsilon} (x,v,t)) 
\;\; \mbox{in}\;\; \Omega\times V\times (0,T) \label{eq11cc}\\
& & g_{\epsilon}(x,v,t)=g_{\epsilon 0}(x,v,t) \;\; \mbox{on}\;\;
\Gamma_0^-\times (0,T) \label{eq13cc}\\
& & g_{\epsilon}(x,v,t)=g_{\epsilon 1}(x,v,t) \;\; \mbox{on}\;\;
\Gamma_1^{-}\times (0,T), \;\; \;\;  \label{eq12cc}\\
& & \nonumber \\
& & g_\epsilon(x,v,0)=g^0_\epsilon(x,v) 
\;\;\mbox{in}\;\;\Omega\times V\label{eq14cc}
\end{eqnarray}

\noindent where all data and the relationships between the various
quantities above were precised in the introduction, we only need
to take $d=1$. We assume that the conservation law (\ref{eq3ncc}) 
is nonlinear in the sense that there exists no interval on which the
flux $A(u)$  is linear, i.e. $A^{\prime \prime} (u)\neq 0$ a.e.
In the full space case i.e. $\Omega=\ir$, the study
of this problem without using compactness arguments (based on $BV$ 
estimates as in Lemma \ref{lem4}) has been done in \cite{bp}.
The authors use compensated compactness, specifically, the Tartar's
div-curl lemma \cite{tartar}. We shall extend this result to the case of
domains with boundaries. We first give a definition of a solution to the
nonlinear conservation laws.

\begin{defi} \label{defi2}
We say that $w\in L^\infty(\Omega \times [0,T])$
is a weak entropic solution of the problem
(\ref{eq3ncc})-(\ref{eq6cc}) if we have

\begin{eqnarray}
& & -\int_{\Omega\times (0,T)}(|w-k| \partial_t
\psi+\mbox{sign}(w-k)(A(w)-A(k))\cdot \nabla_x \psi) \nonumber \\
& &
+\int_{\Gamma_1 \times (0,T)}\psi
\mbox{sign}(w_1-k)((A(w_1)\cdot n)^- - (A(k)\cdot n)^-) \nonumber \\
& & 
+\int_{\Gamma_0^- \times(0,T)} a(v)\cdot n \psi |g_0-\chi_k| 
\le 0 \nonumber \\
& & \nonumber \\
& & \forall \psi \in
C^1_0(\bar{\Omega}\times V \times (0,T)),\;
\psi \ge 0,\;  \forall k \in \ir \nonumber 
\end{eqnarray}

and $w$ satisfies the initial condition

\[
w(x,0)=w^0(x)\;\;   \mbox{in}\;\; \Omega
\]

\end{defi}

The main result of this section is

\begin{teo}\label{teocc}
Assume that the conservation law (\ref{eq3ncc}) is nonlinear (see
above).
Let $g_\epsilon$ be the solution of the corresponding kinetic equation 
(\ref{eq11cc})-(\ref{eq14cc}). Assume that

\begin{eqnarray}
& &
\|g_{\epsilon 0}\|_{L^\infty(\Gamma_0^{-}\times [0,T])}<C_1,\;\;
\|g^0_\epsilon\|_{L^\infty(\Omega\times V)}<C_2,\;\; 
\|g_{\epsilon 1}\|_{L^\infty(\Gamma_1^{-}\times [0,T])}<C_3,\nonumber \\
& &
\|g^0_\epsilon \|_{L^1(\Omega\times V)}<C_4, \;\;
\|a(v)\cdot n g_{\epsilon 0}\|_{L^1(\Gamma_0^{-}\times (0,T))}<C_5,\;\; 
\|a(v)\cdot n g_{\epsilon 1}\|_{L^1(\Gamma_1^{-}\times (0,T))}<C_6  
\nonumber
\end{eqnarray}

\noindent with $C_i,i=1,\cdots,6$ positive constants 
independent of $\epsilon$. 

Assume also that the initial and boundary data $g_{\epsilon
0}$, $g_\epsilon^0$, and $g_{\epsilon 1}$ are compactly supported in
$v\in V$ with
supports included in a fixed compact set independent of
$\epsilon$. Finally assume that as $\epsilon \rightarrow 0$, 

\begin{eqnarray}
& &
\|w_\epsilon(\cdot,0)-w^0(\cdot)\|_
{L^1_{loc}(\Omega)}
=\|\int_V g_\epsilon^0(\cdot,v)-w^0(\cdot)\|_
{L^1_{loc}(\Omega)}
\rightarrow 0
\label{eqini1cc} \\
& & a(v)\cdot n g_{\epsilon 0} \rightarrow a(v)\cdot n g_{0}
\;\;\mbox{strongly in}\;\; L^1(\Gamma_0^- \times (0,T))
\label{dat3cc} \\
& & a(v)\cdot n g_{\epsilon 1} \rightarrow a(v)\cdot n g_{1}=a(v)\cdot
n\chi_{w_1} 
\;\;\mbox{strongly in}\;\; L^1(\Gamma_1^- \times (0,T))
\label{dat4cc} 
\end{eqnarray}

Then $w_\epsilon =\int_V g_\epsilon(x,v,t)dv$ converges strongly in
$L^p(\Omega \times (0,T))$, $p<\infty$, to an entropic solution 
of the nonlinear conservation law (\ref{eq3ncc})-(\ref{eq6cc})
 in the sense of Definition \ref{defi2}.
\end{teo}

\begin{rema} \label{remacc}
1) We observe that under the assumptions of the theorem above,
the conclusions of Lemmas \ref{lem1}, \ref{lem2}, and 
\ref{lemc1} remain valid.

2) Remark \ref{rema2} is also valid for Theorem \ref{teocc}.
\end{rema}

\vskip .5cm
\noindent
{\bf Proof of Theorem \ref{teocc}}
\vskip .5cm

The proof follows the same lines as the one corresponding to the full
space case in \cite{bp}. Thus, proceeding as in \cite{bp}, we obtain

\begin{eqnarray}
& & \overline{\int_V a(v) g_\epsilon dv} =\overline{\int_V a(v) \chi_{w_\epsilon}
dv}=\overline{A(w_\epsilon)} \label{eqacc1} \\
& & 
\overline{A(w_\epsilon)}=A(\overline{w_\epsilon}) \label{eqacc0}
\end{eqnarray}

\noindent for otherwise, $\overline{|w_\epsilon-\overline{w_\epsilon}|}(x,t)=0$,
which in turn yields again (\ref{eqacc0}).
Combining (\ref{eqacc1}) and (\ref{eqacc0}), and passing to the
limit weakly in (\ref{eq11cc}), we obtain

\[
\frac{\partial}{\partial t} \overline{w_\epsilon}+
\frac{\partial}{\partial x} A(\overline{w_\epsilon})=0
\]

Hence a subsequence of $w_\epsilon$ (still denoted $w_\epsilon$)
converges to a weak solution of the conservation law (\ref{eq3ncc}).
Thanks to the nonlinearity of $A(w)$ and equality (\ref{eqacc0}), we can
use Tartar Theorem [\cite{tartar}, Theorem 26] to conclude that 
$w_\epsilon$ strongly converges in $L^p_{loc}(\Omega \times (0,T))$, $1
\le p<\infty$. This combined with the process used to prove Theorem
\ref{teokcl} completes the proof of the theorem.

\section{Conservation laws with source terms}

In this section we introduce the following 
kinetic model with forces 

\begin{eqnarray}
& & [\partial_t+a(v)\cdot \partial_x +S(x,t,v)\cdot \partial_v]
g_{\epsilon}(x,v,t)=\frac{1}{\epsilon}
(\chi_{w_{\epsilon}(x,t)}(v)-g_{\epsilon} (x,v,t)) \nonumber \\
& &
\;\;\;\;\;\;\;\;\;\; \mbox{in}\;\; \Omega\times V\times (0,T) \label{eq11s}\\
& & g_{\epsilon}(x,v,t)=g_{\epsilon 0}(x,v,t) \;\; \mbox{on}\;\;
\Gamma_0^-\times (0,T) \label{eq13s}\\
& & g_{\epsilon}(x,v,t)=g_{\epsilon 1}(x,v,t) \;\; \mbox{on}\;\;
\Gamma_1^{-}\times (0,T), \;\; \;\;  \label{eq12s}\\
& & \nonumber \\
& & g_\epsilon(x,v,0)=g^0_\epsilon(x,v) 
\;\;\mbox{in}\;\;\Omega\times V\label{eq14s}
\end{eqnarray}

\noindent and study its relation to the inhomogeneous scalar conservation laws

\begin{eqnarray}
& & \partial_t w(x,t)+\partial_{x_i} [A_i(w)](x,t) =S(x,t,w) 
\;\; \mbox{in}\;\; \Omega_{g}\times (0,T) \label{eq3ns}\\
& & \mbox{Boundary conditions for $w$ on $\Gamma_0\times (0,T)$
and $\Gamma_1\times (0,T)$} \label{eq4s}\\
& & \nonumber \\
& & w(x,0)=w^0(x) \;\;\mbox{in}\;\; \Omega \label{eq6s}
\end{eqnarray}

\noindent Here, $S(x,t,.)$ is a source term, which is in 
$L^\infty(\Omega \times (0,T);C^1)$ and satisfies $S(,x,t,0)\equiv 0$.
As before $w_\epsilon(x,t)=\int_V g_\epsilon (x,v,t)dv$
and $\chi_w$ is defined by the relation (\ref{eqsig}). 

In the full space case $\Omega=\ir^d$, a brief study of the 
inhomogeneous scalar conservation laws above has been given 
in \cite{bp} in connection with the kinetic model 

\begin{eqnarray}
& & [\partial_t+a(v)\cdot \partial_x]
g_{\epsilon}(x,v,t)=\frac{1}{\epsilon}
(\chi_{w_{\epsilon}(x,t)}(v)-g_{\epsilon} (x,v,t)) +
S^\prime(x,t,v)g_\epsilon(x,v,t)
\nonumber \\
& &
\;\;\;\;\;\;\;\;\;\; \mbox{in}\;\; \Omega\times V\times (0,T) \label{eq11sn}\\
& & g_{\epsilon}(x,v,t)=g_{\epsilon 0}(x,v,t) \;\; \mbox{on}\;\;
\Gamma_0^-\times (0,T) \label{eq13sn}\\
& & g_{\epsilon}(x,v,t)=g_{\epsilon 1}(x,v,t) \;\; \mbox{on}\;\;
\Gamma_1^{-}\times (0,T), \;\; \;\;  \label{eq12sn}\\
& & \nonumber \\
& & g_\epsilon(x,v,0)=g^0_\epsilon(x,v) 
\;\;\mbox{in}\;\;\Omega\times V\label{eq14sn}
\end{eqnarray}

As compared with the kinetic model (\ref{eq11sn})-(\ref{eq14sn})
proposed in \cite{bp},
our kinetic model (\ref{eq11s})-(\ref{eq14s}) is more appropriate to
describe the physics at the microscopic
level, which yields the conservation laws (\ref{eq3ns})-(\ref{eq6s}) at
the macroscopic level as the miscropscopic scale tends to 0.
Its analysis does not require additional assymptions on the source 
terms as in \cite{bp}. We shall clarify this later.

Since our kinetic model is new, we shall also indicate how our analysis 
extend to the full space case i.e. $\Omega=\ir^d$.
 
We begin with an existence and uniqueness result for the kinetic model.

\begin{teo}
Assume that

\[
g^0_\epsilon \in L^1(\Omega\times V), \;\;
a(v)\cdot n g_{\epsilon 1}\in 
L^1(\Gamma_1^{-}\times (0,T)), \;\;
a(v)\cdot n g_{\epsilon 0} \in L^1(\Gamma_0^{-}\times (0,T))
\]

\noindent
Then the kinetic model (\ref{eq11s})-(\ref{eq14s}) has a unique solution in
$L^\infty([0,T];L^1(\Omega \times V))$. Moreover, $g_\epsilon$ 
satisfies the integral representation

\begin{eqnarray*}
& & \mbox{In} \;\; \Omega_0\;\; \\ 
& &
\;\;g_\epsilon(x,v,t)=
g_\epsilon((0,x_\star -\frac{x_1}{a_1(v)}a_\star(v)),
v-\frac{x_1}{a_1(v)}S(x,t,v),t-\frac{x_1}{a_1(v)})
\mbox{exp}(-\frac{x_1}{\epsilon a_1(v)})+
\\ & &
+\frac{1}{\epsilon} \int_{t-\frac{x_1}{a_1(v)}}^t
\mbox{exp}((s-t)/\epsilon)\chi_{w_\epsilon(x(s),s)}(v(s))ds \\
& & \\
& & \mbox{In} \;\; \Omega_{01}\\
& &
\;\; g_\epsilon(x,v,t)=g^0_\epsilon(x-a(v)t,v-tS(x,t,v))
\mbox{exp}(-t/\epsilon)
+\frac{1}{\epsilon} \int_{0}^t
\mbox{exp}((s-t)/\epsilon)\chi_{w_\epsilon(x(s),s)}(v(s))ds \\
& & \\
& & \mbox{In} \;\; \Omega_1 \;\; \\ 
& &
\;\; g_\epsilon(x,v,t)=g_{\epsilon
1}((1,x_\star+\frac{1-x_1}{a_1(v)}a_\star(v)),v+\frac{1-x_1}{a_1(v)}
S(x,t,v),t-\frac{x_1-1}{a_1(v)})
\mbox{exp}(\frac{1-x_1}{\epsilon a_1(v)}) 
\\ & &
+\frac{1}{\epsilon} \int_{t-\frac{x_1-1}{a_1(v)}}^t
\mbox{exp}((s-t)/\epsilon)\chi_{w_\epsilon(x(s),s)}(v(s))ds \\
\end{eqnarray*}

\noindent where $x(s)=x+(s-t)a(v)$, $x=(x_1,x_\star)$, 
$a(v)=(a_1(v),a_\star(v))$, and $v(s)=v+(s-t)S(x,t,v)$.

Finally, Let $g_\epsilon$ and
$G_\epsilon$ be two solutions of (\ref{eq11})-(\ref{eq14})
with corresponding densities $w_\epsilon(x,t)=\int_V g_\epsilon(x,v,t)dv$ and
$W_\epsilon(x,t)=\int_V G_\epsilon(x,v,t)dv$; and let 
$g^0_\epsilon,\;g_{\epsilon 0},\;g_{\epsilon 1}$ 
resp. $G^0_\epsilon,\;g_{\epsilon 0},\;G_{\epsilon 1}$ 
denote the corresponding data. 
Let 

$$S^\prime_\infty(t)=\{ \mbox{max}_{x,v} S^\prime(x,t,v): \;v\in
\mbox{supp}_{v} g_\epsilon(x,v,t) \cup \mbox{supp}_{v} G_\epsilon(x,v,t)\}$$
We have

\begin{eqnarray}
& &
\|g_\epsilon-G_\epsilon\|_{L^\infty([0,T];L^1(\Omega\times V))}
\nonumber \\
&\le& 
\mbox{exp}(\int_0^T |S^\prime_\infty(s)|ds) \nonumber \\
& &  [
\|g_\epsilon^0-G_\epsilon^0\|_{L^1(\Omega \times V)}+
\|a(v)\cdot n (g_{\epsilon 0}-G_{\epsilon
0})\|_{L^1(\Gamma_0^{-}\times(0,T))}+
\|a(v)\cdot n (g_{\epsilon 1}-G_{\epsilon
1})\|_{L^1(\Gamma_1^{-}\times (0,T))}]
\label{estcont1} \\
& & \nonumber \\
& & \nonumber \\
& &
\|g_\epsilon-G_\epsilon\|_{L^\infty([0,T];L^1(\Omega\times V))}+
\|a(v)\cdot n (g_{\epsilon 0}-G_{\epsilon
0})\|_{L^1(\Gamma_0^{+}\times(0,T))} \nonumber \\
& & +
\|a(v)\cdot n (g_{\epsilon 1}-G_{\epsilon 1})\|_{L^1(\Gamma_1^{+}\times(0,T))}
\nonumber \\
&\le& 
[1+\int_0^t |S^\prime_\infty(s)|\mbox{exp}(\int_0^\sigma 
|S^\prime_\infty(\sigma)|d\sigma)ds]  [
\|g_\epsilon^0-G_\epsilon^0\|_{L^1(\Omega \times V)}+
\nonumber \\
& & 
\|a(v)\cdot n (g_{\epsilon 0}-G_{\epsilon
0})\|_{L^1(\Gamma_0^{-}\times(0,T))}+
\|a(v)\cdot n (g_{\epsilon 1}-G_{\epsilon
1})\|_{L^1(\Gamma_1^{-}\times (0,T))}]
\label{estcont2}
\end{eqnarray}

\end{teo}

The proof of this theorem follows by arguing along the lines of 
the proof of 
Theorem \ref{teok}, with obvious
modification to account for the source term. We only point out here how 
to integrate by part in the term $S \partial_v g_\epsilon$. Let
$\varphi$ be as in the proof of Theorem \ref{teok}. Let $\eta \in
C_0^\infty(V)$ satisfy $0\le \eta \le 1$, $\eta \equiv 1$ on $[-1,1]$,
and $\mbox{supp} \eta \subset [-2,2]$. Let $\eta_n=\eta(v/n)$.
After multiplying the equation for $g_\epsilon$ by $\varphi \eta_n$,
the contribution of the source term is

\begin{eqnarray}
& &
\int_{\Omega \times V \times (0,t)}S(x,t,v) \partial_v g_\epsilon
\varphi \eta_n \nonumber \\
&=& -\int_{\Omega \times V \times (0,t)}g_\epsilon 
\partial_v S(x,t,v) \varphi \eta_n-g_\epsilon S(x,t,v)\partial_v \varphi
\eta_n -g_\epsilon S(x,t,v)\varphi \partial_v \eta_n 
\end{eqnarray}

After passing to the limit as $n \rightarrow \infty$, the right hand
side converges to 
$$-\int_{\Omega \times V \times (0,t)}g_\epsilon 
\partial_v S(x,t,v) \varphi -g_\epsilon S(x,t,v)\partial_v \varphi
$$

We also pass to the limit as $n \rightarrow \infty$ in the other terms.
The rest of the proof proceeds as in the proof of Theorem \ref{teok}
with appropriate modifications due to the source term.

We shall give below an entropy inequality for the solution of the kinetic
problem. This is stated in the following theorem.

\begin{teo} \label{teoents}
The solution to the kinetic problem satisfies the
relation

\begin{eqnarray}
& &
-\int_{\Omega \times V\times (0,T)}
(\partial_t+a(v)\cdot \partial_x)(\psi)|g_\epsilon-\chi_k|
+\int_{\Gamma_0^-\times (0,T)} a(v)\cdot n 
\psi|g_{\epsilon 0}-\chi_k|+ \nonumber \\
& &
\int_{\Gamma_1^{-}\times (0,T)} a(v)\cdot n 
\psi|g_{\epsilon 1}-\chi_k|
\le \int_{\Omega \times V\times (0,T)} g_\epsilon \psi \partial_v S 
\mbox{sign}(g_\epsilon -\chi_k)
\label{ke4s}
\end{eqnarray}

\begin{eqnarray*}
& & \forall \psi \in C_0^1(\bar{\Omega}\times V\times (0,T)), \;
\psi \ge 0,\; \forall k\in \ir 
\end{eqnarray*}

\end{teo}

Before we state our main convergence results, we 
shall give below a definition of a solution to
the conservation laws with source term 
(\ref{eq3ns})-(\ref{eq6s}). This definition 
selects a physically correct solution to this problem.

\begin{defi} \label{defi1s}
We say that $w\in BV_{loc}(\Omega \times(0,T)) \cap 
L^\infty(\Omega \times [0,T])$
is a weak entropic solution of the problem
(\ref{eq3ns})-(\ref{eq6s}) if we have

\begin{eqnarray}
& & -\int_{\Omega\times (0,T)}(|w-k| \partial_t
\psi+\mbox{sign}(w-k)(A(w)-A(k))\cdot \nabla_x \psi) \nonumber \\
& &
+\int_{\Gamma_1 \times (0,T)}\psi
\mbox{sign}(w_1-k)((A(w_1)\cdot n)^- - (A(k)\cdot n)^-) \nonumber \\
& & 
+\int_{\Gamma_0^- \times(0,T)} a(v)\cdot n \psi |g_0-\chi_k| 
\le \int_{\Omega\times (0,T)} \psi S(x,t,w) \mbox{sign}(w-k)      
 \nonumber \\
& & \nonumber \\
& & \forall \psi \in
C^1_0(\bar{\Omega}\times V \times (0,T)),\;
\psi \ge 0,\;  \forall k \in \ir \nonumber 
\end{eqnarray}

and $w$ satisfies the initial condition

\[
w(x,0)=w^0(x)\;\;   \mbox{in}\;\; \Omega
\]

\end{defi}

We mention here that the kinetic entropy relations given in \cite{bp}
on page 516, Formula (5.5) for the kinetic model 
(\ref{eq11sn})-(\ref{eq14sn})
and their corresponding 
macroscopic ``continuum limit'' entropy inequality given at the end of
page 516 in \cite{bp}
for the conservation laws with source terms (\ref{eq3ns})-(\ref{eq6s})
are not correct.

Next we shall state the main convergence results about the kinetic
distributions and their moments for the source case.

\begin{teo}\label{teocls0}
Assume that 

\begin{eqnarray*}
& &
\|g_{\epsilon 0}\|_{L^\infty(\Gamma_0^{-}\times [0,T])}<C_1,\;\;
\|g^0_\epsilon\|_{L^\infty(\Omega\times V)}<C_2,\;\; 
\|g_{\epsilon 1}\|_{L^\infty(\Gamma_1^{-}\times [0,T])}<C_3,\nonumber \\
& &
\|g^0_\epsilon \|_{L^1(\Omega\times V)}<C_4, \;\;
\|a(v)\cdot n g_{\epsilon 0}\|_{L^1(\Gamma_0^{-}\times (0,T))}<C_5,\;\; 
\|a(v)\cdot n g_{\epsilon 1}\|_{L^1(\Gamma_1^{-}\times (0,T))}<C_6  
\nonumber \\
& &
\|g_\epsilon^0\|_{L^1(V;BV_{loc}(\Omega))} <C_{7} \nonumber
\end{eqnarray*}

\noindent with $C_i,i=1,\cdots,7$ positive constants 
independent of $\epsilon$. 

Assume also that the initial and boundary data $f_{\epsilon
0}$, $g_\epsilon^0$, and $g_{\epsilon 1}$ are compactly supported in
$v\in V$ with
supports included in a fixed compact set independent of
$\epsilon$. Finally assume that as $\epsilon \rightarrow 0$, 

\begin{eqnarray}
& &
\|w_\epsilon(\cdot,0)-w^0(\cdot)\|_
{L^1_{loc}(\Omega)}
=\|\int_V g_\epsilon^0(\cdot,v)-w^0(\cdot)\|_
{L^1_{loc}(\Omega)}
\rightarrow 0
\label{eqini1s} \\
& & a(v)\cdot n g_{\epsilon 0} \rightarrow a(v)\cdot n g_{0}
\;\;\mbox{strongly in}\;\; L^1(\Gamma_0^- \times (0,T))
\label{dat3s} \\
& & a(v)\cdot n g_{\epsilon 1} \rightarrow a(v)\cdot n g_{1}=a(v)\cdot
n\chi_{w_1} 
\;\;\mbox{strongly in}\;\; L^1(\Gamma_1^- \times (0,T))
\label{dat4s} 
\end{eqnarray}

\noindent
Then $w_\epsilon$ converges strongly in $L^1(\Omega \times V \times
(0,T))$,as $\epsilon$ goes to $0$, to an entropic solution of the problem
(\ref{eq3n})-(\ref{eq6}) in the sense of Definition \ref{defi1}.
\end{teo}

The theorem above does not provide a strong convergence uniform in
$\epsilon$ and time of the density
distribution to the equilibrium distribution. This is due to the
presence of initial layers and the lack of the control of the velocity
variation of the density distribution. Under the present assumptions 
(assumptions of Theorem \ref{teocls0}) only a uniform control of 
the spatial variation on the microscopic scale and a uniform control of 
the temporal variation only at the macroscopic level are allowed (consult 
 Lemma \ref{lem4} part 1) and 2) and the remark after the proof of
Theorem in the sourceless case). The uniform control of the temporal
variation of the kinetic distribution can be achieved only if we can 
control uniformly, in addition to the spatial variation, the
the velocity variation and the initial temporal variation of the kinetic
distribution. That is, we have to prepare the initial data 
so that $\frac{\partial}{\partial t}
g_\epsilon$ is uniformly bounded in $\epsilon$ and $t$,
in particular at $t=0$, and $g_\epsilon^0$ is uniformly bounded in
$BV(V;L^1_{loc}(\Omega))$. 
We therefore assume that the kinetic initial data satisfies \cite{bp}

\begin{eqnarray*}
& & \|g_\epsilon^0(\cdot,\cdot)-\chi_{w^0(\cdot)}(\cdot)\|_
{L^1_{loc}(\Omega\times L^1(V))}
\rightarrow_{\epsilon \rightarrow 0} 0 \\
& &
\|g_\epsilon^0\|_{L^1_{loc}(\Omega;B(V))}  <C
\end{eqnarray*}

Under the new additional assumptions, we obtain the following uniform 
in $\epsilon$ and time  convergence of the kinetic ditribution to an 
equilibrium distribution.

\begin{teo}\label{teocls2}
Assume that 

\begin{eqnarray*}
& &
\|g_{\epsilon 0}\|_{L^\infty(\Gamma_0^{-}\times [0,T])}<C_1,\;\;
\|g^0_\epsilon\|_{L^\infty(\Omega\times V)}<C_2,\;\; 
\|g_{\epsilon 1}\|_{L^\infty(\Gamma_1^{-}\times [0,T])}<C_3,\nonumber \\
& &
\|g^0_\epsilon \|_{L^1(\Omega\times V)}<C_4, \;\;
\|a(v)\cdot n g_{\epsilon 0}\|_{L^1(\Gamma_0^{-}\times (0,T))}<C_5,\;\; 
\|a(v)\cdot n g_{\epsilon 1}\|_{L^1(\Gamma_1^{-}\times (0,T))}<C_6  
\nonumber \\
& &
\|g_\epsilon^0\|_{L^1(V;BV_{loc}(\Omega))} <C_{7},\;\;
\|g_\epsilon^0\|_{L^1_{loc}(\Omega;BV(V))} <C_{8}
 \nonumber
\end{eqnarray*}

\noindent with $C_i,i=1,\cdots,8$ positive constants 
independent of $\epsilon$. 

Assume also that the initial and boundary data $f_{\epsilon
0}$, $g_\epsilon^0$, and $g_{\epsilon 1}$ are compactly supported in
$v\in V$ with
supports included in a fixed compact set independent of
$\epsilon$. Finally assume that as $\epsilon \rightarrow 0$, 

\begin{eqnarray}
& &
\|g_\epsilon^0(\cdot,\cdot)-\chi_{w^0(\cdot)}(\cdot)\|_
{L^1_{loc}(\Omega\times L^1(V))}
\rightarrow_{\epsilon \rightarrow 0} 0
\label{eqini2s} \\
& & a(v)\cdot n g_{\epsilon 0} \rightarrow a(v)\cdot n g_{0}
\;\;\mbox{strongly in}\;\; L^1(\Gamma_0^- \times (0,T))
\label{dat3ns} \\
& & a(v)\cdot n g_{\epsilon 1} \rightarrow a(v)\cdot n g_{1}=a(v)\cdot
n\chi_{w_1} 
\;\;\mbox{strongly in}\;\; L^1(\Gamma_1^- \times (0,T))
\label{dat4ns} 
\end{eqnarray}

\noindent
Then $g_\epsilon$ converges strongly in $L^\infty([0,T];L^1(\Omega
\times V))$, as $\epsilon$ goes to $0$, to $\chi_w$ and
$w$ is an entropic solution of the problem
(\ref{eq3n})-(\ref{eq6}) in the sense of Definition \ref{defi1}.
\end{teo}

\begin{rema} 1) Remark \ref{rema2} is also valid for Theorems
\ref{teocls0} and \ref{teocls2}. 

2) Notice that Theorems \ref{teocls0} and \ref{teocls2} are also valid
for the simpler case of full space $\Omega=\ir^d$ with appropriate
modifications. We shall compare below our results for the full space case to
those of \cite{bp}.
For our generalized kinetic model the corresponding theorem to 
Theorem \ref{teocls0} for the full space case 
is obtained under no additional assumptions on the data or source terms.
The analysis in \cite{bp} required the additional assumption that
the source terms are in $BV(\Omega)$. However, to obtain  the uniform in
$\epsilon$ and time convergence of the density distribution to 
an equilibrium distribution (the corresponding theorem 
to Theorem \ref{teocls2} for the full space case), we had to assume an 
additional assumption that the initial ditribution 
$g_\epsilon^0$ is uniformly bounded in $L^1_{loc}(\Omega;BV(V))$.
As a result in our case the existence theory for conservation laws with
source terms is obtained under no additional assumptions on the source 
terms as opposed to the existence theory given in \cite{bp} which
required the additional assumption that the source terms are $BV$.
Thus our theory is more general.
\end{rema}

To prove these theorems we argue along the lines of the proof of Theorem
\ref{teokcl} for the sourceless case, with appropriate
modifications due to the source term. We shall therefore state without
proofs the corresponding lemmas with the necessary modifications caused 
by the presence of the source term.

We begin with $L^\infty$ estimates.

\begin{lem}\label{lem1s}
Assume that 
\[
\|g_{\epsilon 0}\|_{L^\infty(\Gamma_0^{-}\times [0,T])} <C_1,\; 
\|g^0_\epsilon\|_{L^\infty(\Omega\times V)}<C_2, \; 
\|g_{\epsilon 1}\|_{L^\infty(\Gamma_1^{-}\times [0,T])}<C_3
\]

\noindent with $C_1,C_2,$ and $C_3$ positive constants independent of 
$\epsilon$.  Then 
$g_\epsilon$ is uniformly bounded in
$L^\infty(\Omega \times V\times[0,T])$.
Moreover we have

\begin{eqnarray*}
\|g_\epsilon\|_{\infty}  
&\le&  
[\mbox{max}(\|g_{\epsilon 0}\|_{L^\infty(\Gamma_0^{-}\times [0,T])},
\|g^0_\epsilon\|_{L^\infty(\Omega\times V)},
\|g_{\epsilon 1}\|_{L^\infty(\Gamma_1^{-}\times [0,T])})+1]
\mbox{exp}(\int_0^T |S^\prime_\infty(\tau)|d\tau)
\end{eqnarray*}

\noindent Here $$S^\prime_\infty(t)=
\{ \mbox{max}_{x,v} S^\prime(x,t,v): \;v\in
\mbox{supp}_{v} g_\epsilon(x,v,t) \}$$
\end{lem}

\begin{lem}\label{lem2s}

\noindent
Assume that

\begin{eqnarray*}
& & \|a(v)\cdot n g_{\epsilon 0}\|_{L^1(\Gamma_0^{-}\times (0,T))} <C_1,\;\;
\|g^0_\epsilon\|_{L^1(\Omega\times V)}<C_2, \\
& &
\|a(v)\cdot n g_{\epsilon 1}\|_{L^1(\Gamma_1^{-}\times (0,T))}<C_3  
\end{eqnarray*}

\noindent
with $C_1,C_2,$ and $C_3$ positive constants independent of $\epsilon$.
Then $g_\epsilon$ is uniformly bounded in
$L^\infty([0,T];L^1(\Omega\times V))$ and
$w_\epsilon$ is uniformly bounded  in 
$L^\infty([0,T];L^1(\Omega))$. Moreover, we have

\begin{eqnarray*}
\|w_\epsilon\|_{L^\infty([0,T];L^1(\Omega))} 
&\le& 
\|g_\epsilon\|_{L^\infty([0,T];L^1(\Omega\times V))}
\nonumber \\
&\le & \mbox{exp}(\int_0^T |S^\prime_\infty(\tau)|d\tau)
[\|a(v)\cdot n g_{\epsilon 0}\|_{L^1(\Gamma_0^{-}\times (0,T))}+
\|a(v)\cdot n g_{\epsilon 1}\|_{L^1(\Gamma_1^{-}\times (0,T))}+
\nonumber \\
& &
+\|g_\epsilon^0\|_{L^1(\Omega\times V)}]
\end{eqnarray*}

\end{lem}

\begin{lem} \label{lemc1s}
Assume that 
\begin{eqnarray*}
& &
\|g_{\epsilon 0}\|_{L^\infty(\Gamma_0^{-}\times [0,T])}<C_1,\;\;
\|g^0_\epsilon\|_{L^\infty(\Omega\times V)}<C_2,\\
& &
\|g_{\epsilon 1}\|_{L^\infty(\Gamma_1^{-}\times [0,T])}<C_3
\end{eqnarray*}

\noindent with $C_1,C_2,$ and $C_3$ positive constants 
independent of $\epsilon$. 
Assume also that the initial and boundary data 
$g_\epsilon^0$, $g_{\epsilon 0}$, and $g_{\epsilon 1}$ are compactly 
supported in $v\in V$ with supports included in a fixed compact set 
independent of $\epsilon$. 
Then 

\noindent (i) $w_\epsilon$ is uniformly bounded in
$L^\infty(\Omega\times [0,T])$. 

\noindent (ii)
$g_\epsilon$ remains compactly supported in $v\in V$
with support included in a fixed compact set independent of
$\epsilon$.

\noindent (iii)
The speed of propagation $a(v)$ is finite.
\end{lem}

\begin{lem}\label{lem4s}
Assume that 

\begin{eqnarray}
& &
\|g_{\epsilon 0}\|_{L^\infty(\Gamma_0^{-}\times [0,T])}<C_1,\;\;
\|g^0_\epsilon\|_{L^\infty(\Omega\times V)}<C_2,\;\; 
\|g_{\epsilon 1}\|_{L^\infty(\Gamma_1^{-}\times [0,T])}<C_3,\nonumber \\
& &
\|g^0_\epsilon \|_{L^1(\Omega\times V)}<C_4,\;\; 
\|a(v)\cdot n g_{\epsilon 0}\|_{L^1(\Gamma_0^{-}\times (0,T))}<C_5,\;\; 
\|a(v)\cdot n g_{\epsilon 1}\|_{L^1(\Gamma_1^{-}\times (0,T))}<C_6  
\nonumber \\
& &
\|g_\epsilon^0\|_{L^1(V;BV_{loc}(\Omega))} <C_{7},\nonumber 
\end{eqnarray}

\noindent with $C_i,i=1,\cdots,7$ positive constants 
independent of $\epsilon$. 
Assume also that the initial and boundary data $f_{\epsilon
0}$, $g_\epsilon^0$, and $g_{\epsilon 1}$ are compactly supported in
$v\in V$ with
supports included in a fixed compact set independent of
$\epsilon$. 

\noindent Then 

1) $g_\epsilon(\cdot,\cdot,t)$ and $w_\epsilon(\cdot,t)$, 
$t\in [0,T]$ are uniformly bounded 
in $BV_{loc}(\Omega\times L^1(V))$
and $BV_{loc}(\Omega)$ respectively.
More precisely, if
$U$ and $O$ are open bounded subsets of $\Omega$ such that
$\bar{U}\subset O \subset \bar{O} \subset \Omega$, we have
for $i=1,\cdots,d$

\[
\int_{U\times V} |\tau^i_h g_\epsilon-g_\epsilon| \le 
\mbox{exp}(\int_0^t |S^\prime_\infty(s)| ds)
\int_{O\times V} |\tau^i_h g_\epsilon^0-g_\epsilon^0|
\]

2)
$w_\epsilon$ is time Lipschitz continuous in $L^1_{loc}(\Omega)$
uniformly in $\epsilon$; i.e. for any open bounded subset $U$ of
$\Omega$ with $\bar{U}\subset \Omega$, we have

\begin{eqnarray}
& &
\|w_\epsilon(\cdot,t_2)-w_\epsilon(\cdot,t_1)\|_{L^1(U)} \nonumber \\
& <&
(a_\infty\|g_\epsilon\|_{L^\infty([0,T];BV(U\times L^1(V)))} 
+\|\partial_v S\|_{L^\infty(\Omega \times
[0,T])}\|g_\epsilon\|_{L^\infty(\Omega \times V\times [0,T])})
(t_2-t_1) \nonumber \\
&<& C(t_2-t_1),\nonumber \\
& & \forall \;0\le t_1<t_2\le T  \label{eqbvloct1s}
\end{eqnarray}

\noindent where $C$ is a constant depending on $U$ but is
independent of $\epsilon$ and $a_\infty$ is introduced in the proof of
Lemma \ref{lemc1} above.

3) Under the additional assumptions

\begin{eqnarray*}
& &
\|g_\epsilon^0(\cdot,\cdot)-\chi_{w^0(\cdot)}(\cdot)\|_
{L^1_{loc}(\Omega\times L^1(V))}
\rightarrow_{\epsilon \rightarrow 0} 0
\nonumber \\
& &
\|g_\epsilon^0\|_{L^1_{loc}(\Omega; BV(V))} <C_{8},\nonumber 
\end{eqnarray*}

\noindent $g_\epsilon(\cdot,\cdot,t)$, $t\in [0,T]$ is uniformly bounded 
in $BV_{loc}(V;L^1_{loc}(\Omega))$. Moreover, we can estimate the error 
between the kinetic solution and exact entropy solution as follows

\begin{eqnarray}
\|g_\epsilon -
\chi_{w_\epsilon}\|_{L^\infty([0,T];L^1_{loc}(\Omega\times L^1(V)))}
&\le& \epsilon a_\infty
\|g_\epsilon^0(x,v)\|_{BV_{loc}(\Omega\times L^1(V))}
\nonumber \\
& &
+\epsilon a_\infty
\|g_\epsilon(x,v,t)\|_{L^\infty([0,T];BV_{loc}(\Omega\times L^1(V))))}
\nonumber \\
& & +2 \|g_\epsilon^0(x,v)-\chi_{w^0(x)}\|_
{L^1_{loc}(\Omega\times L^1(V))}\nonumber \\
& & +\epsilon \mbox{max}_v \|S\|_{L^\infty(\Omega \times
[0,T])} \nonumber \\
& & (\|g_\epsilon(x,v,t=0)\|_{BV(V\times L^1(O))}+
\|g_\epsilon(x,v,t)\|_{BV(V\times L^1(O))}) \nonumber \\
& &
    \rightarrow_{\epsilon \rightarrow 0} 0 
\label{eqbvloct5s}
\end{eqnarray}

4) The function $w_\epsilon$
is uniformly bounded in $BV_{loc}(\Omega\times(0,T))$.

\end{lem}

\bibliographystyle{amsplain}

\end{document}